\documentstyle[a4,11pt,epsfig]{article}
\newcommand \beq{\begin{eqnarray}}
\newcommand \eeq{\end{eqnarray}}
\def\leftrightarrowfill{$ \mathord\leftarrow \mkern-6mu \cleaders
\hbox{$\mkern-2mu \mathord- \mkern-2mu$}\hfill \mkern-6mu \mathord\rightarrow$}
\def\overleftrightarrow#1{ \vbox{\ialign{##\crcr \leftrightarrowfill\crcr
\noalign{\kern-1pt\nointerlineskip}
$\hfil\displaystyle{#1}\hfil$\crcr}}}
\def\sqr#1#2{{\vcenter{\vbox{\hrule height.#2pt \hbox {\vrule width.#2pt
height#1pt \kern#1pt \vrule width.#2pt} \hrule height.#2pt }}}}

\begin{document}
\title{\sf Propagation of neutral mesons in asymmetric nuclear matter}
\author{\sf L. Mornas \\ 
\\
{\small{\it Departamento de F\'{\i}sica, Universidad de Oviedo, 
Avda Calvo Sotelo s/n, E-33007 Oviedo, Spain}} }
\maketitle
\par\noindent {\small keywords: relativistic RPA, 
proton-neutron asymmetry, in medium properties of mesons}

\begin{abstract} 
We calculate dispersion relations and propagators for the $\sigma$, $\omega$,  
$\rho$ and $\delta$ mesons in relativistic nuclear matter with a
proton-neutron asymmetry. In addition to the $\sigma$-$\omega$ and 
$\delta$-$\rho$ mixings already present in symmetric matter, mixing 
occur between all components of the $\sigma$-$\omega$-$\delta$-$\rho$ 
system when the proton fraction differs from 1/2. 
\end{abstract}

%%%%%%%%%%%%%%%%%%%%%%%%%%  1 - Introduction     %%%%%%%%%%%%%%%%%%%%%%%%%%

\section{Introduction}

This work extends previous results on the dispersion relation of mesons
in relativistic, dense, hot nuclear matter, to systems where there exists
a proton-neutron asymmetry. We will be working in a model of the 
quantum hadrodynamics type (QHD) \cite{SW86}, where the nucleon-nucleon 
interaction is mediated by the exchange of $\omega$, $\sigma$, $\pi$ 
... {\it etc.} mesons. In this work, the dispersion relations of
the mesons are calculated by performing a linear response analysis
around an asymmetric Hartree gound state. The propagators are then obtained
by inverting the dispersion relations. The resulting expression
corresponds to the random phase approximation. 

A new feature of the asymmetric matter is that new meson mixing
channels open. A standard case of meson mixing is that occurring
for the $\sigma$-$\omega$ pair, where one meson is converted
into the other by desintegration and recombination of a virtual
NN loop. Moreover, the isoscalar mesons can couple with the isovector 
ones whenever the distribution function of proton and neutrons differ.
This has been calculated earlier for the $\omega$-$\rho$ pair 
(see {\it e.g.} \cite{DMDRK97,BF98}) in relation with the 
interpretation of dilepton production in heavy ion collisions 
\cite{Bratkovskaya,Sarkar98}. The $\delta$-$\rho$ 
mixing was recently investigated by \cite{TDMG00} in 
symmetric matter as a further mechanism contributing to  
dilepton production, and also by \cite{japaneseDR}
and \cite{M01}.
We present here a derivation and systematical study of the 
dispersion relation and propagators of the coupled neutral 
meson system  $\omega$-$\sigma$-$\rho^0$-$\delta^0$. The charged mesons
will be studied elsewhere.

Our calculations may apply to the physics of neutron stars or of heavy 
ion collisions. 

Nuclear physics in terrestrial conditions is for the most part 
well described by the $N=Z$ hypothesis. Nevertheless, new data begin 
to be available with sizeable deviations from symmetry, by performing 
experiments with nuclei at the border of the stability lines. Indeed, 
with the planned construction of radioactive beam facilities in several 
countries, exciting new possibilities will open of investigating the 
equation of state and dynamical properties of asymmetric nuclear matter. 
Such experiments will hopefully be able to shed some light on 
the behavior of the asymmetry energy at high density \cite{LiKo}.

Among nuclear physics applications, one can think for example of observing 
the excitation of collective modes in neutron rich nuclei \cite{Klaus}.
Another application concerns the description of the isospin dynamics in  
heavy ion collisions at intermediate energies. In particular, one of 
the accessible experimental observables, the balance energy, defined as 
the beam energy at which transverse collective flow disappears, is valued 
as providing a clean way  to estimate the influence of medium efects on 
the nucleon-nucleon cross section \cite{KWB82}. Previous work \cite{DM98a} 
has shown that, besides short range effects described the Brueckner 
approximation, one cannot neglect the screening of the nuclear interaction 
arising from the exchange of dressed mesons. An interesting phenomenon
would occur if a zero sound mode is excited, since it would appear as a 
pole in the meson propagator and a corresponding resonant behavior of 
the cross section \cite{DM98b}. This analysis can be extended in order 
to include the description of isospin effects \cite{Liu2001,HCM99,DB98}.

\vskip 0.5cm

Large neutron to proton ratio is the rule in astrophysical 
situations, where the matter is in $\beta$ equilibrium. In cold neutron
stars, the proton fraction is about 10 \% or lower. In supernovae
explosions and hot proneutron stars, the proton fraction is higher,
about 30 to 40 \%, but deviations from the symmetric nuclear matter 
cannot be neglected.

In the context of the physics of supernova collapse and the early 
stages of proto neutron star cooling, the screening of the 
nucleon-nucleon interaction has consequences on the 
neutrino opacities \cite{Lattimer,YT00,NHV99,MP01}. 
In dense and hot matter, the neutrino-nucleon scattering rate is modified 
by the correlations which affect the nucleon current. In the random phase 
approximation, the neutrino-nucleon differential scattering cross section
is given by
\beq
{d \sigma \over d E_\nu d \Omega} = -{G_F^2 \over 32 \pi^3}
{E_{\nu '} \over E_\nu} {1 \over 1-e^{-z}}
{\cal I}m (\widetilde\Pi_{WW}^{(R)\, \mu\nu} L_{\mu\nu})     
\eeq
where $\Pi_{WW}^{(R)\, \mu\nu}$ is the retarded polarization 
obeying the Dyson equation  
\beq
\widetilde\Pi_{WW}^{\mu\nu}=\Pi_{WW}^{\mu\nu} +  \Pi_{WS}^{\mu\alpha}
\widetilde D_{SS\, \alpha\beta} \Pi_{SW}^{\beta\nu}
\label{Dyson}
\eeq 
$W$ and $S$ represent vertices for the weak and strong couplings 
respectively
and $\widetilde D_{SS\, \alpha\beta}$ is the dressed meson propagator.
This propagator involves the meson dispersion relation in the 
denominator. The magnitude of the scattering rate therefore depends 
crucially on whether or not zero-sound modes turns up in the meson 
propagation within the integration range. As an example,
the Skyrme calculation of \cite{NHV99}, which present a spin zero sound 
mode, predict an enhancement of neutrino opacities, whereas the relativistic
approach of \cite{Lattimer,YT00,MP01}, where this mode is absent, 
predict a reduction of the opacities. Further study is necessary.

The calculation of \cite{Lattimer,YT00} did not take into account the 
full picture of meson propagation in asymmetric matter. We have applied 
\cite{MP01} the propagators derived in the present work to the 
determination of the neutrino-nucleon scattering rate, and find
that the reduction effected by RPA correlations is less efficient 
when all  mixing channels are taken into consideration.

%%%%%%%%%%%%%%%%%%%%%%   2  - The Model   %%%%%%%%%%%%%%%%%%%%%%%%%%

\section{The Model}

\subsection{Lagrangian density}

We will use a  version of the quantum hadrodynamics model developped 
by Walecka and coworkers in which the nucleon-nucleon interaction is 
mediated the six mesons taken into account by the Bonn potential model, 
{\it i.e} $ \sigma$, $\omega$, $\pi$, $\rho$, $\pi$, $\eta$ and $\delta$ 
mesons. The latter two mesons are often neglected in mean-field type 
calculations by arguing that they lead to comparatively small corrections 
to the nucleon-nucleon interaction potential in the vacuum \cite{M89,M87}. 
Nevertheless, the $\delta$ meson can play a significant role in the 
context of dense asymmetric matter, since the value of the $\delta$ 
mean field controls the mass difference between proton and neutron. 
This field modifies the behaviour of the neutron star equation of state 
\cite{Kutschera}. The value of the $\delta NN$ coupling has
a strong influence on the proton fraction of matter in $\beta$ 
equilibrium. Another justification is the fact that an extrapolation 
of Dirac-Brueckner-Hartree-Fock calculations to asymmetric matter was 
shown to be equivalent to a density-dependent mean field theory with 
a delta meson \cite{dJL98,FLW95,SST97}.

The $\pi$ and $\eta$ will mix with each other, and separate from the 
$\omega$-$\sigma$-$\rho$-$\delta$ sector. We will not study the
$\pi$-$\eta$ system here, since our main purpose was in fact to 
apply the results derived in this work to the calculation of 
the RPA reduction to the neutrino-nucleon scattering rate,
where the pion does not contribute \cite{MP01}. The calculation
of the mixed $\pi$-$\eta$ dispersion relations would not 
anyway present any particular difficulty, using the same 
method as applied here to derive the equations which describe the
$\sigma$-$\delta$ subsystem. 

We start from the Lagrangian density
\beq 
{\cal L} &=& 
\hat{\overline \psi} \left[ i \gamma .{\overleftrightarrow{\partial}
\over 2} -m +g_\sigma \hat \sigma -g_\omega \gamma^\mu \hat \omega_\mu 
+ g_\delta \hat{\vec \delta}.\vec \tau -g_\rho \gamma^\mu \hat{\vec \rho}_\mu 
. \vec \tau + {i f_\rho \over 2 m} \sigma^{\mu\nu} \partial_\nu 
\hat{\vec \rho}_\mu .\vec \tau \right] \hat\psi \nonumber \\
& & -{1 \over 4} \hat F_{\mu\nu} \hat F^{\mu\nu} + {1 \over 2} m_\omega^2 
\hat\omega_\mu   \hat\omega^\mu  +{1 \over 2} \partial_\mu \hat\sigma 
\partial^\mu \hat\sigma -{1 \over 2} m_\sigma^2  \hat\sigma^2 
-{1 \over 3} b m \hat \sigma^3 - {1 \over 4} c \hat\sigma^4 \nonumber \\
& & +{1 \over 2} \hat{\vec \delta}.\hat{\vec \delta} 
-{1 \over 2} m_\delta^2 \hat\delta^2 -{1 \over 4} \hat{\vec R}_{\mu\nu}.
\hat{\vec R}^{\mu\nu} +{1 \over 2} \hat{\vec \rho}_\mu.\hat{\vec \rho}^\mu
\label{Lagrangian}
\eeq
The hat on the Wigner operator and meson fields means that they are
operators. The pion does not appear in this Lagrangian because it will 
not be investigated further in this work. Terms such as $\hat\sigma 
\hat\pi^2$ or $\hat{\vec \rho}^\mu . (\partial_\mu \hat{\vec \pi} 
\times \hat{\vec \pi} )$ also do not appear, beacause we decided to 
focus in this work on meson mixing originating from NN loops, and will 
not treat meson loops. A more realistic treatment should include 
them as well.

This Lagrangian is non renormalizable {\it stricto sensu} due to the 
derivative couplings of the $\rho$ meson (and $\pi$ if included
with a pseudovector coupling). Nevertheless, this coupling is 
necessary for a good description of experimental data. In fact, 
renormalizability is not a necessary requirement for effective 
theories \cite{Furnstahl}. Anyway, a renormalization procedure 
should be applied at the given level of approximation in order 
to avoid pathological behavior of the dispersion relation. To 
this end, we will add counterterms to the Lagrangian. 
A discussion of this issue is given in section \S \ref{polariz} and 
Appendix B.

\subsection{Relativistic Hartree equilibrium}

The approach taken by \cite{JDAetal} is applied here. The main 
steps leading to the dispersion relations are only briefly recorded; 
the reader is referred for details to \cite{JDAetal}. It consists of
writing a kinetic equation for the Wigner operator, and perform a linear
response analysis of perturbations of the system around the 
equilibrium as obtained in the Hartree approximation.
We first define the Wigner function
\beq
\hat F(x,p)= \int {d^4\, R \over (2 \pi)^2}\ \exp(-i p.R)\ 
\hat{\overline \psi} \left( x+{R \over 2} \right) \otimes
\hat\psi \left( x-{R \over 2} \right)
\eeq
In the Hartree approximation, we suppose that the meson field operators 
$\hat\sigma$, $\hat\omega$ ... can be replaced by their classical 
expectation values $\sigma_H$, $\omega_H$ ... and that the equilibrium 
is homogeneous. We will moreover assume that there does not exist
charged condensates (only the third component of $\vec\delta_H$ 
and $\vec\rho_H$
are non zero), so that the proton and neutron states remain well
defined and the Wigner function is diagonal in isospin space. 
The nucleon fields are then plane wave solutions of a Dirac equation 
with effective momenta and masses. The Wigner function 
is obtained by introducing these solutions in the definition and 
taking the statistical average. It is given by
\beq
F_H(p) &=& \left[ \matrix{ \displaystyle{1\over (2 \pi)^3} \delta( P_p^2 - M_p^2)
(\gamma.P_p + M_p) f_p (P_p) & 0 \cr
0 & \displaystyle{1\over (2 \pi)^3} \delta( P_n^2 - M_n^2)(\gamma.P_n + M_n) 
f_n (P_n) \cr} \right] \nonumber \\
& &  \\
{\rm with} & & P_p^\mu= p^\mu -g_\omega \omega_H^\mu - g_\rho \rho_{H}^\mu
    \qquad;  \qquad M_p = m -g_\sigma \sigma_H -g_\delta \delta_H \\
& & P_n^\mu= p^\mu -g_\omega \omega_H^\mu + g_\rho \rho_{H}^\mu
    \qquad;  \qquad M_n = m -g_\sigma \sigma_H +g_\delta \delta_H  \\
\noalign{\smallskip}
& &  \omega_H^\mu= {g_\omega \over m_\omega^2} \int d^4\, p\ {\rm Tr}
\left[ \gamma^\mu F_H(p) \right] \ \quad ; \qquad \sigma_H= {g_\sigma \over
m_\sigma^2}  \int d^4\, p\ {\rm Tr} \left[ F_H(p) \right] \\
& & {\vec \rho}_H^\mu= {g_\rho \over m_\rho^2} \int d^4\, p\ {\rm Tr}
\left[ \gamma^\mu \vec \tau F_H(p) \right] \quad ; \qquad {\vec \delta}_H= 
{g_\delta \over m_\delta^2}  \int d^4\, p\ {\rm Tr} \left[ \vec \tau F_H(p) 
\right] \\
\noalign{\smallskip}
& & f_i(p) = \left[ {\theta (p_0) \over e^{\beta \left( P_i.u -
\mu_i \right)} +1} + {\theta (-p_0) \over e^{-\beta \left( P_i.u - \mu_i \right)} 
+1} -\theta (-p_0) \right] \quad , \quad i \in \{ p,n \}
\label{flucvac}
\eeq
The middle two lines define the effective masses and momenta of the
neutron and proton in terms of the classical fields. They are self consistent 
relations which can be solved for a given total density, proton/neutron 
asymmetry and temperature.  

The equation of state is obtained by calculating the energy momentum 
tensor and conserved baryon current
\beq
J^\mu &=& n_B\, u^\mu \quad ; \quad n_B= n_n + n_p \nonumber \\
      &=& \int d^4\, p\ {\rm Tr} \left[ \gamma^\mu F_H (p) \right] 
\label{eosJ} \\
T^{\mu\nu} &=& (\rho +P)\,  u^\mu u^\nu -P\, g^{\mu\nu} \nonumber \\
           &=& \int d^4\, p\ p^\nu\ {\rm Tr} \left[ \gamma^\mu F_H (p) \right] 
\nonumber \\
           & & +{1 \over 2} g^{\mu\nu} \left( m_\sigma^2 \sigma_H^2 
+ {1 \over 3} b m \sigma_H^3 + {1 \over 4} c \sigma_H^4- m_\omega^2 \omega_H^2
+ m_\delta^2 \delta_{H}^2 - m_\rho^2 \rho_{H}^2
\right) 
\label{eosT}
\eeq
For further details on the equation of state, the reader may wish to
consult the work of Kubis and Kutschera \cite{Kutschera}.

The proton fraction is defined as
\beq
Y_p= {n_p \over n_n + n_p}
\eeq
It can be given as an input parameter or be
determined from the $\beta$ equilibrium condition. The first 
case would be relevant for heavy ion collisions, where the 
asymmetry is determined  by the choice of the nuclear isotopes
participating to the reaction. In neutron stars and supernovae
on the other hand, the timescales are long enough for the chemical 
equilibrium to establish through inverse $\beta$ decay 
\beq
p + e^- \rightarrow n + \nu
\eeq
yielding a condition on the chemical potentials which determines $Y_p$
\beq
\hat \mu = \mu_n - \mu_p = \mu_e - \mu_\nu
\eeq
%For example, at vanishing temperature,
%$\hat \mu = \sqrt{ \k_{Fn}^2 +M_n^2} - \sqrt{ \k_{Fp}^2 +M_p^2} 
% -2 (g_\rho/m_\rho)^2 (n_p -n_n)$ with $n_i=k_{Fi}^3/(3 \pi^2)$.
In cold neutron stars, the neutrinos escape freely from the system, and we
can moreover set the chemical potential of the neutrinos to zero. 
The proton fraction is of the order of 0.1 depending on the density
which is considered. In the last stages of supernova collapse and early 
phase of protoneutron star cooling on the other hand, the matter is 
opaque to neutrinos and $\mu_\nu$ is determined by dynamical equations 
governing the lepton fraction $Y_L=Y_\nu + Y_e$. Charge conservation 
imposes $Y_e=Y_p$. For a typical value $Y_L$=0.4, we obtain
$\mu_\nu = \left( 6 \pi^2 n_B Y_\nu\right)^{1/3}$ of the order of
200 -- 250 MeV and $Y_p \sim $ 0.3 -- 0.36.
 
An example of the thermodynamics which are obtained from this model 
is shown on Fig. 1, where the effective masses and proton fraction are
represented as a function of density for neutrino-free matter in 
beta equilibrium. It was calculated with parameter set (\ref{fiteos}) 
which fits the saturation point. Due to the coupling to the $\delta$ field, 
the proton effective mass is higher than that of the neutron in this model.
For the same value of the asymmetry energy at saturation, larger proton
fractions are obtained at high density with a finite value of the
coupling to the delta field, {\it e.g.} $g_\delta=5$, than 
$g_\delta=0$. 

%%%%%%%%%%%%%%%%%%%%%%   3  - Dispersion relations    %%%%%%%%%%%%%%%%%%%%%%%%%%

\section{Dispersion relations}

\subsection{Linear response}

The Hartree solution was a particular approximation to the general
evolution equations
\beq
& & \left[ i {\gamma.\partial \over 2} + (\gamma.p -m) \right] \hat F (x,p) =
- \int \, {d^4\, R \over (2 \pi^4)}\, d^4\, \xi \ e^{-i (p-\xi).R }
\ \hat \Phi(x-{R \over 2}) \hat F(x,\xi) 
\label{kinetic} \\
& &  {\rm with} \quad \hat\Phi(x)=
g_\sigma\, \hat \sigma(x) + g_\delta\,  \hat{\vec\delta}(x) . \vec \tau 
- g_\omega\, \gamma^\mu \hat\omega_\mu(x)  \nonumber \\
& & \qquad \qquad \quad  -g_\rho\, \gamma^\mu \hat{\vec\rho_\mu}(x) . \vec \tau 
+ i {f_\rho \over 2 m} \sigma^{\mu\nu} \partial_\nu \hat{\vec\rho_\mu}(x) 
.\vec \tau 
\label{interaction} \\
\noalign{\smallskip}
& & \partial_\mu \hat F^{\mu\nu} + m_\omega^2 \hat \omega^\nu = g_\omega 
\int d^4\, p\ {\rm Tr} \left[ \gamma^\nu \hat F(x,p) \right] \\
& & \left[ \partial_\mu \partial^\mu + m_\sigma^2 \right] \hat \sigma 
+ b m \hat\sigma^2 +c \hat \sigma^3 = g_\sigma \int d^4\, p\ {\rm Tr}
\left[ \hat F(x,p) \right] \\
& & \partial_\mu \hat{\vec R}^{\mu\nu} + m_\rho^2 \hat{\vec\rho}^\nu 
=  g_\rho  \int d^4\, p\ {\rm Tr} \left[ \gamma^\nu \vec \tau 
\hat F(x,p) \right] 
-i {f_\rho \over 2 m} \partial_\mu  \int d^4\, p\ {\rm Tr} \left[ 
\sigma^{\mu\nu} \vec \tau \hat F(x,p) \right] \\
& & \left[ \partial_\mu \partial^\mu + m_\delta^2 \right] \hat{\vec \delta} 
= g_\delta \int d^4\, p\ {\rm Tr} \left[ \tau \hat F(x,p) \right] 
\eeq
The hat on the Wigner operator and meson fields means that they are
operators. The physical quantities are extracted by taking statistical 
averages. Up to now, these equations are exact. A first approximation 
will be made by neglecting correlations, {\it i.e.} replacing the 
statistical average of products of meson operators with the
Wigner operator, by the product of statistical averages: 
$< \hat\Phi \hat F>\ \rightarrow\ <\hat\Phi> <\hat F>$.

Let us now consider a small perturbation around the Hartree equilibrium.
We will develop $<\hat F(x,p)>=F_H(p) + F_1(x,p)$, $<\hat \sigma(x)>
=\sigma_H + \sigma_1(x)$ ... {\it etc}. The  Hartree equilibrium 
is obtained at zeroth order. At first order, we obtain equations for 
the perturbation which, after performing a Fourier transformation, 
are easily solved.
The dispersion relations of the mesons are obtained by replacing
the solution for the perturbation to the Wigner function 
\beq
F_1 &=& G(p-{q \over 2})\ {\cal S}(q)\ F_H (p+{q \over 2})
+ F_H(p-{q \over 2})\ {\cal S}(q)\ G (p+{q \over 2}) \\
{\rm with} & &\!\!\!\!\! {\cal S}(q) = -g_\sigma \sigma_1(q) 
+ g_\omega \gamma^\mu \omega_{1 \mu}(q) - g_\delta \vec \delta_1(q).\vec \tau
+g_\rho \gamma^\mu \vec \rho_{1 \mu}(q). \vec \tau + {f \rho \over 2 m} 
\sigma^{\mu\nu} q_\nu \vec \rho_{1 \mu}.\vec \tau \\
{\rm and}  & &\!\!\!\!  G(p) = \left[ \matrix{ \displaystyle{\gamma.P_p +M_p 
\over P_p^2 -M_p^2}  & 0 \cr 0 & \displaystyle{\gamma.P_n +M_n \over 
P_n^2 -M_n^2} } \right]
\label{propagator}
\eeq
in the linearized, Fourier-transformed field equations for the
mesons. For example, the dispersion relation for the sigma meson reads
\beq
(-q^2 + m_\sigma^2 +2\, b\, m\, \sigma_H +3\, c\, \sigma_H^2)\ \sigma_1 
     &=& g_\sigma \int d^4\, p\ {\rm Tr} [ F_1(p,q) ]  \nonumber \\
                             &=& \Pi_{\sigma\sigma}\ \sigma_1 + 
\Pi_{\sigma\omega}^\mu\ \omega_{1 \mu} + \Pi_{\sigma\delta}^a\ \delta_{1a} 
+ \Pi_{\sigma\rho}^{a\mu}\ \rho_{1a \mu} 
\eeq
This defines the polarizations.
The full set of dispersion relations is given in matricial form 
in \S \ref{disprel}. 
The index $a$ is an isospin index. When taking the traces,
it can be seen that the dispersion relations of the neutral mesons
$\sigma$, $\omega$, $\rho^0$, $\delta^0$, $\pi^0$ decouple from those
of the charged mesons $\rho^\pm$, $\delta^\pm$, $\pi^\pm$. In this 
paper we limit ourselves to the dispersion relations of the neutral
mesons; the case of the charged mesons will be the subject of future 
work.

The nonlinear $\sigma^3$, $\sigma^4$ couplings
contribute terms on the left hand side which we can absorb in
a definition 
\beq
M_\sigma^2 = m_\sigma^2  +2\, b\, m\, \sigma_H +3\, c\, 
\sigma_H^2.
\label{Msig}
\eeq
We point out again that there are no meson loops (such as would be 
contributed {\it e.g.} by a $\pi$-$\pi$ loop insertion in the $\sigma$ 
propagator) at this level of approximation in our formalism. They were 
discarded when we made the assumption $<\hat\Phi \hat F> \ <\hat\Phi>
<\hat F>$. One could obtain them by restoring meson-meson correlations,
or insert such loops by hand basing ourselves on the diagrammatic approach .
This will not be done in this work, since our first aim was to study
meson mixing originating in NN loops.

\subsection{Polarizations}
\label{polariz}

Let us continue on the example of the $\sigma$ meson. Explicitely,
the $\Pi_{\sigma\sigma}^{(i)}$ polarization is given by
\beq
\Pi_{\sigma\sigma}=\Pi_{\sigma\sigma}^{(p)} + \Pi_{\sigma\sigma}^{(n)}
\nonumber \\
\Pi_{\sigma\sigma}^{(i)} &=& - \int d^4\, p\ {\rm Tr} \left[
 g_\sigma G \left(P_i +{q \over 2} \right) g_\sigma  F_H 
\left(P_i -{q \over 2} \right) + ( G \leftrightarrow F_H)  \right] 
\nonumber
\eeq
At zero temperature, the result obtained through the application of the linear
response analysis coincides with the one-loop approximation. 
The other polarizations are given by similar equations. Taking the traces 
and integrating over the angles, we finally arrive at the formulaes 
given in the Apendix. Some remarks are in order about the imaginary part
at finite temperature and about the contribution of vacuum fluctuations.

\vskip 0.5cm

We will have to choose a prescription to go around the poles of
the propagator Eq. (\ref{propagator}). 
In \cite{JDAetal}, the prescription was chosen as follows
\beq
\Pi_{\sigma\sigma}^{(i)} &=& - g_\sigma^2 \int d^4\, p\ {\rm Tr} \left[
\{ g_\sigma \} \left\{ \gamma.\left( P_i -{q \over 2} \right) +M_i \right\}
\{ g_\sigma \}\left\{ \gamma.\left( P_i +{q \over 2} \right) +M_i \right\}
\right] \times \nonumber \\
& & \times \left[ {\delta \left( \left( P_i +\displaystyle{q \over 2} 
\right)^2 -M_i^2 \right) \ f_i\left(P_i+\displaystyle{q \over 2} \right)  
\over (2 \pi)^3\ \left[ \left( P_i -\displaystyle{q \over 2} \right)^2 
-M_i^2 +i \epsilon \right] } + 
 {\delta \left( \left( P_i -\displaystyle{q \over 2} \right)^2 
-M_i^2 \right) \ f_i\left(P_i-\displaystyle{q \over 2} \right)  \over
(2 \pi)^3\ \left[ \left( P_i +\displaystyle{q \over 2} \right)^2 -M_i^2 
-i \epsilon \right] } \right] \nonumber \\
&=& 4 g_\sigma^2 \int d^4\, p \left[ p^2 -{q^2 \over 4}
+ M_i^2 \right] \left\{ {\varphi_i (p + q/2) - \varphi_i (p-q/2) 
\over 2 p.q -i \epsilon} \right\} 
\label{epsilonJDA}
\eeq
with $\varphi_i(p) = \displaystyle{1 \over (2 \pi)^2} \delta(P_i^2-M_i^2) 
f_i(p) $

At finite temperature, the real time Green's function formalism 
\cite{SMS89-FS91-KS856} defines several polarizations depending on the 
position of the time arguments on the Keldysh contour. Their real parts 
coincide, the imaginary parts are related to each other by relations 
such as ${\cal I}m\ \Pi^R (w,k) =\tanh \left( {\beta \omega \over 2} 
\right) \ {\cal I}m\ \Pi^{11} (w,k)$.
It can be verified that the usual retarded polarizations 
\cite{SMS89-FS91-KS856} can be obtained by taking $\pm\, i\ \epsilon\ 
{\rm sign(p_0)}$ instead of $\pm\, i\ \epsilon$ in the above expression
(\ref{epsilonJDA}). The imaginary parts given in the Appendix as well 
as those used in the calculation of the propagators will be
those of the retarded polarizations.

\vskip 0.5cm

The last term $\theta(-p_0)$ in the expression of the Wigner function 
Eq. (\ref{flucvac}) describes the contribution of the vacuum. When taken 
under the integration sign in the expression of the polarizations, 
it yields a divergent term.
Some renormalization procedure has to be applied in order to remove the 
divergence.
It is not sufficient to ignore this term (by performing a normal ordering),
because the subtracted term depends on the effective masses and therefore
on the thermodynamical conditions. In the following we will apply dimensional
regularization and counterterm subtraction as described {\it e.g.} in
\cite{C77,JDAetal,MGP98}. What is new with respect to these references is 
that, due to proton-neutron asymmetry, the contributions of the mesons cannot
be renormalized independently.

The mixed polarizations $\Pi_{\sigma\rho}^\mu$ and $\Pi_{\delta\omega}^\mu$
are both proportional to $\eta^\mu = u^\mu - (q.u / q^2) q^\mu$. They must 
vanish in vacuum, since the hydrodynamical velocity quadrivector is
undefined in this case. Therefore, we can treat the renormalization of the
$\sigma$-$\delta$ subsystem independently from that of the $\omega$-$\rho$
subsystem. 
We review in this section the main steps of the renormalization of the 
$\sigma$-$\delta$ subsystem and give the expressions of all renormalized 
vacuum contributions in Appendix B.

After performing the dimensional regularization, the divergent
contributions are extracted in the form $(M_p^2 + M_n^2)/ \epsilon$
and $q^2/ \epsilon$ in $\Pi_{\sigma\sigma}^{\rm vac}$ or
$\Pi_{\delta\delta}^{\rm vac}$, and in the form $(M_p^2 - M_n^2) /
\epsilon$ in $\Pi_{\sigma\delta}^{\rm vac}$ or $\Pi_{\delta\sigma}^{\rm
vac}$.
The fact that $M_n^2 + M_p^2= 2 ( m -g_\sigma \sigma)^2 + 2 g_\delta^2 
\delta^2 $ and $M_p^2 -M_n^2 = 2 (m-g_\sigma \sigma) g_\delta \delta$ 
dictates the form of the counterterm Lagrangian one has to introduce,
in order to neutralize these divergences.
\beq
{\cal L}_{\rm CT} &=& Z_\sigma\ \partial_\mu \sigma \partial^\mu \sigma  
+ {1 \over 2} A_1\ \sigma^2 + {1 \over 3} A_2\ \sigma^3 
+ {1 \over 4} A_3\ \sigma^4 \nonumber \\
&& +Z_\delta\ \partial_\mu \delta \partial^\mu \delta 
+ B_1\ \delta + B_2\ \sigma \delta^2 + B_3\ \sigma^2 \delta^2 +C\ \delta^4
\eeq
Each of the couplings of the counterterm Lagrangian ($Z_i$, $A_i$, 
$B_i$, $C$) is split ($A_1=$ infinite $+ a_1$, {\it etc}) in an 
infinite part, which is chosen so as to cancel the divergence, 
and a residual finite part ($z_i$, $a_i$, $b_i$, $c$), which has to be 
determined by imposing some physical conditions that the 
renormalized polarizations should verify. 

There are several possible choices, each of which defines a 
renormalization procedure. One possibility is to impose that the 
polarization and its first derivatives with respect to $\sigma$ and 
$\delta$ vanish at some point, for example on the mass shell $q^2=m_\sigma^2$
(as {\it e.g.} in the work of Diaz Alonso {\it et al.} \cite{JDAetal}) 
or at $q^2=0$ (as in the work of Kurasawa and Suzuki \cite{KS88}). 
Another possibility is to require that the initial structure of the expression 
in terms of the effective masses is preserved (The procedures of 
{\it e.g.} Hatsuda {\it et al.} \cite{SH94} or Sarkar {\it et al} 
\cite{Sarkar98} do have this property, as well as  ``scheme 3''  
of \cite{MGP98}.) Here we will work with the scheme of
Kurasawa and Suzuki (called ``scheme A'' in the following)
and scheme 3 of \cite{MGP98} called here ``scheme B'').

As was already pointed out in \cite{MGP98}, the final result, 
and in particular the behavior of the effective meson masses as 
a function of density, is very sensitive to the choice of 
renormalization procedure. We must consider this as an 
unsolved problem for the moment. There is some hope that one could 
solve it by applying ``naturalness'' arguments \cite{Furnstahl}
and considerations on the symmetries of the  underlying 
more fundamental theory, of which the present effective one represents 
a low-energy approximation. 

Actually, the asymmetric case offers us an example of how symmetry 
arguments do impose some further restrictions on the choice
of the subtraction procedure. As a matter of fact, the mixing
between the $\sigma$ and $\delta$ mesons requires that the {\it same}
counterterms should simulaneously cancel divergences in various 
polarizations. For example, the $B_3$ is used to regulate the 
$\delta^2/\epsilon$ divergence in $\Pi_{\sigma\sigma}$, the
$\sigma^2/\epsilon$ divergence in $\Pi_{\delta\delta}$ and
the $\sigma \delta/\epsilon$  divergences in $\Pi_{\sigma\delta}$
and $\Pi_{\delta\sigma}$. The $B_2$ regulates divergences
$\delta/\epsilon$ in  $\Pi_{\sigma\delta}$ and $\Pi_{\delta\sigma}$,
 and $\sigma/\epsilon$ in $\Pi_{\delta\delta}$.
The compatibility will impose additional relations between the finite
$b_2$, $b_3$, {\it etc.} constants. It was found that
the renormalization scheme used by \cite{JDAetal} must be discarded 
on this basis, since the additional constraints necessary for the 
compatibility of the counterterms are not fulfilled. 
Both scheme A and B used in this paper fulfill the compatibility 
relations. Scheme B preserves the structure of the expression
as a function of the effective masses $M_p$, $M_n$ but
gives after renormalization  $\Pi_{\sigma\delta} 
\not=\Pi_{\delta\sigma}$, whereas these two quantities 
were given by the same expression before renormalization.
For scheme A the  reverse occurs, we have  $\Pi_{\sigma\delta} 
=\Pi_{\delta\sigma}$ after renormalization but the 
structure in terms of $M_p$, $M_n$ is not maintained.

Further details are given in Appendix B.

\subsection{Dispersion relations}
\label{disprel}

The equations for the dispersion relations can be cast in matricial form:
\beq
[{\cal D}][\Phi_1]=0 \ ,\qquad
{\cal D}= \left[ \matrix{ D_\sigma & D_{\sigma\omega}^\mu & D_{\sigma\delta}
& D_{\sigma\rho}^\mu \cr
\noalign{\smallskip}
D_{\omega\sigma}^\nu & D_{\omega\omega}^{\mu\nu} & D_{\omega\delta}^\nu &
D_{\omega\rho}^{\mu\nu} \cr
\noalign{\smallskip}
D_{\delta\sigma} & D_{\delta\omega}^\nu & D_{\delta\delta} & 
D_{\delta\rho}^\nu \cr
\noalign{\smallskip}
D_{\rho\sigma}^\nu & D_{\rho\omega}^{\mu\nu} & D_{\rho\delta}^\nu &
D_{\rho\rho}^{\mu\nu} \cr } \right]\ ,\qquad
\Phi_1=\left[ \matrix{ \sigma_1 \cr \omega_{1\mu} \cr \delta_1 \cr
\rho_{1 \mu}} \right] 
\eeq
In a referential where the momentum transfer is equal to $q^\mu= 
(\omega,0,0,k)$, the ${\cal D}$ matrix has the structure
\begin{table}[htbp]
\begin{center}
\begin{tabular}{|c|cccc|c|cccc|}
q & r & 0 & 0 & s & q1 & r1 & 0 & 0 & s1 \\
\hline
r & a & 0 & 0 & b & t1 & v & 0 & 0  & z \\ 
0 & 0 & c & 0 & 0 & 0 & 0 & x & 0 & 0 \\  
0 & 0 & 0 & c & 0 & 0 & 0 & 0 & x & 0 \\
s & b & 0 & 0 & e & u1 & z & 0 & 0 & y  \\
\hline
q1 & t1 & 0 & 0 & u1 & d & t & 0 & 0 &  u  \\
\hline
r1 & v & 0 & 0 & z & t & f & 0 & 0 & j \\
0 & 0 & x & 0 & 0 & 0 & 0 & h & 0 & 0 \\
0 & 0 & 0 & x & 0 & 0 & 0 & 0 & h & 0 \\
s1 & z & 0 & 0 & y & u & j & 0 & 0 & i \\
\end{tabular} 
\end{center}
\end{table}
\par\noindent
with the following definitions of $a$, $b$, $c$, $d$ ... {\it etc}.
\vfill
\newpage

\begin{table}[htbp]
\begin{center}
\begin{tabular}{lcl}
{\rm q}  = $q^2-m_\sigma^2 +\Pi_\sigma$   & \phantom{blank} &
      {\rm q1} = $\Pi_{\sigma\delta}$ \\
{\rm r} = $-k^2/q^2 \ \Pi_{\omega\sigma}$   & \phantom{blank} &
      {\rm r1} = $-k^2/q^2 \ \Pi_{\sigma\delta}$   \\
{\rm s} = $-\omega k /q^2 \Pi_{\omega\sigma}$ & \phantom{blank} &
      {\rm s1} = $-\omega k/q^2 \ \Pi_{\sigma\delta}$  \\
{\rm a} = $k^2 +m_\omega^2 +k^2/q^2 \ \Pi_{\omega L}$  & \phantom{blank} &
      {\rm f} = $k^2 +m_\rho^2 +k^2/q^2 \ \Pi_{\rho L}$  \\
{\rm b} = $\omega k + \omega k/q^2 \ \Pi_{\omega L}$ & \phantom{blank} & 
      {\rm j} = $\omega k + \omega k/q^2 \ \Pi_{\rho L}$ \\
{\rm c} = $q^2-m_\omega^2 + \ \Pi_{\omega T}$  & \phantom{blank} & 
      {\rm k} = $q^2 -m_\rho^2 +\Pi_{\rho T}$ \\
{\rm e} = $\omega^2 -m_\omega^2 +\omega^2/q^2 \ \Pi_{\omega L}$ & 
      \phantom{blank} &  
      {\rm i} = $\omega^2 -m_\rho^2 +\omega^2/q^2 \ \Pi_{\rho L}$ \\
{\rm d} = $q^2-m_{\delta}^2 +\Pi_\delta$ & \phantom{blank} &
      {\rm v} = $k^2/q^2 \ \Pi_{\omega \rho L}$ \\
{\rm t} = $-k^2/q^2 \ \Pi_{\delta\rho}$   & \phantom{blank} &
       {\rm z} = $\omega k/q^2 \ \Pi_{\omega \rho L}$ \\
{\rm u} = $-\omega k/q^2 \ \Pi_{\delta\rho}$  & \phantom{blank}& 
      {\rm x} = $\Pi_{\omega \rho T}$  \\
{\rm t1 } = $-k^2/q^2 \ \Pi_{\omega \delta}$ & \phantom{blank} &
      {\rm y}=  $\omega^2/q^2 \ \Pi_{\omega \rho L}$  \\
{\rm u1} = $-\omega k/q^2 \ \Pi_{\omega\delta}$ & \phantom{blank} &  
\end{tabular}
\end{center}
\end{table}

\par\noindent
In writing the preceding equations, we used the decomposition of the
polarizations on a basis of orthogonal tensors
\beq
\Pi_{\omega}^{\mu\nu} &=&-\Pi_{\omega\, T}\ T^{\mu\nu}-\Pi_{\omega\, L}\ 
L^{\mu\nu} \quad , \qquad
\Pi_{\rho}^{\mu\nu}=-\Pi_{\rho\, T}\ T^{\mu\nu}-\Pi_{\rho\, L}\ L^{\mu\nu}
\nonumber \\
\Pi_{\rho\omega}^{\mu\nu}&=&\Pi_{\omega\rho}^{\mu\nu}=-\Pi_{\rho\omega\, T} 
\ T^{\mu\nu}-\Pi_{\rho\omega\, L}\ L^{\mu\nu} \nonumber \\
\Pi_{\sigma\omega}^{\mu}&=&\Pi_{\omega\sigma}^{\mu}
=\Pi_{\sigma\omega}\ \eta^{\mu}\quad , \qquad
\Pi_{\delta\rho}^{\mu}=\Pi_{\rho\delta}^{\mu}
=\Pi_{\delta\rho}\ \eta^{\mu} \nonumber \\
\Pi_{\sigma\rho}^{\mu}&=&\Pi_{\rho\sigma}^{\mu}=\Pi_{\sigma\rho}\ \eta^{\mu}
\quad , \qquad \Pi_{\delta\omega}^{\mu}=\Pi_{\omega\delta}^{\mu}
=\Pi_{\delta\omega}\ \eta^{\mu} \nonumber 
\eeq
with
\beq
L^{\mu\nu} &=& {\eta^\mu \eta^\nu \over \eta^2} \quad ; \quad
\eta^\mu = u^\mu - {q.u \over q^2} q^\mu \nonumber \\
T^{\mu\nu} &=& g^{\mu\nu} - {\eta^\mu \eta^\nu \over \eta^2} -{q^\mu
q^\nu \over q^2} \quad ; \quad Q^{\mu\nu}={q^\mu q^\nu \over q^2} 
 \nonumber 
\eeq

If the non linear couplings $b,c$ are non zero, one should replace
everywhere $m_\sigma^2$ by $M_\sigma^2$ as defined in Eq. (\ref{Msig}).
Taking the determinant of this matrix, we obtain the dispersion 
relation. It is of the form $det({\cal D}) = m_\rho^2\, m_\omega^2\, 
 \delta_T^2\, \delta_L =0$, the transverse propagation modes being given by
\beq
\delta_T= (q^2-m_\rho^2 +\Pi_{\rho T})(q^2-m_\omega^2+\Pi_{\omega T})
-\Pi_{\rho\omega T}^2 =0
\eeq
and the longitudinal mode by
\beq
\delta_L &=& \left[ (q^2-m_\sigma^2+\Pi_\sigma)
  (q^2-m_\delta^2+\ \Pi_\delta) -\Pi_{\sigma\delta}^2 \right] 
  \left[ (q^2-m_\rho^2 +\Pi_{\rho L}) (q^2-m_\omega^2+\Pi_{\omega L})
  -\Pi_{\rho\omega L}^2 \right] \nonumber \\
         & &  + (q^2-m_\omega^2+\Pi_{\omega L})
                (q^2-m_\sigma^2+\Pi_\sigma)\, \Pi_{\delta\rho}^2\, \eta^2 
        +(q^2-m_\rho^2 +\Pi_{\rho L}) (q^2-m_\sigma^2+\Pi_\sigma)\, 
                \Pi_{\delta\omega }^2\, \eta^2  \nonumber \\
         & & + (q^2-m_\omega^2+\Pi_{\omega L})(q^2-m_\delta^2+\Pi_\delta) 
              \, \Pi_{\sigma\rho}^2\, \eta^2 
             + (q^2-m_\rho^2 +\Pi_{\rho L})(q^2-m_\delta^2+\Pi_\delta) 
              \, \Pi_{\sigma\omega}^2\, \eta^2  \nonumber \\
         & & -2\, \Pi_{\rho\omega L}\, \left[ (q^2-m_\sigma^2+\Pi_\sigma)
             \,\Pi_{\delta\rho}\, \Pi_{\delta\omega }\, \eta^2 +
              (q^2-m_\delta^2+\Pi_\delta)\, \Pi_{\delta\omega}\, 
             \Pi_{\sigma\omega}\,  \eta^2 \right] \nonumber \\
         & & -2\, \Pi_{\sigma\delta}\, \left[ (q^2-m_\omega^2+\Pi_{\omega L})
             \, \Pi_{\delta\rho}\, \Pi_{\sigma\rho}\, \eta^2 
             + (q^2-m_\rho^2+\Pi_{\rho L})\, \Pi_{\delta\omega}\,  
             \Pi_{\sigma\omega}\, \eta^2 \right] \nonumber \\
         & & -2\, \Pi_{\sigma\delta}\, \Pi_{\rho\omega L}\, (\Pi_{\delta\rho} 
             \, \Pi_{\sigma\omega}\,  \eta^2 +\Pi_{\delta\omega }\, 
             \Pi_{\sigma\rho}\,  \eta^2 )  + \eta^4\, \left(
             \Pi_{\delta\omega }\, \Pi_{\sigma\rho}  - \Pi_{\delta\rho}\, 
             \Pi_{\sigma\omega} \right)^2 =0.
\eeq

%%%%%%%%%%%%%%%%%%%%%%%%%%%   4 - Propagators    %%%%%%%%%%%%%%%%%%%%%%%%%%%%%

\section{Propagators}

The propagators are obtained by inverting the dispersion relation
${\cal G}={\cal D}^{-1}$. The propagator matrix has the same structure 
as the dispersion matrix. 
\beq
{\cal G}= \left[ \matrix{ G^\sigma & G^{\sigma\omega}_\mu & G^{\sigma\delta}
& G^{\sigma\rho}_\mu \cr
\noalign{\smallskip}
G^{\omega\sigma}_\nu & G^{\omega\omega}_{\mu\nu} & G^{\omega\delta}_\nu &
G^{\omega\rho}_{\mu\nu} \cr
\noalign{\smallskip}
G^{\delta\sigma} & G^{\delta\omega}_\nu & G^{\delta\delta} & 
G^{\delta\rho}_\nu \cr
\noalign{\smallskip}
G^{\rho\sigma}_\nu & G^{\rho\omega}_{\mu\nu} & G^{\rho\delta}_\nu &
G^{\rho\rho}_{\mu\nu} \cr } \right] \nonumber
\eeq
The scalars isoscalar $G_\sigma$ and isovector $G_{\delta}$ propagators 
are given by
\beq
G_\sigma &=& {1 \over \delta_L} \left\{ 
    (q^2 -m_\omega^2 +\Pi_{\omega L} )\, \Pi_{\delta\rho}^2\, \eta^2 
  + (q^2 -m_\rho^2 +\Pi_{\rho L} )\, \Pi_{\delta\omega}^2\, \eta^2 
  - 2\, \Pi_{\delta\rho}\, \Pi_{\delta\omega}\, \Pi_{\rho\omega L}\, \eta^2  
  \right. \nonumber \\
  & & \left. \ \ \ \ \ + (q^2 -m_\delta^2 +\Pi_\delta)\left[ (q^2 -m_\omega^2 
    +\Pi_{\omega L} ) (q^2 -m_\rho^2 +\Pi_{\rho L}) -\Pi_{\rho\omega L}^2 
    \right] \right\} \\
G_\delta &=& {1 \over \delta_L} \left\{ 
     (q^2 -m_\omega^2 +\Pi_{\omega L})\, \Pi_{\sigma\rho}^2\, \eta^2 + 
     (q^2 -m_\rho^2 + \Pi_{\rho L})\, \Pi_{\sigma\omega}^2\, \eta^2 
     -2\, \Pi_{\sigma\rho}\, \Pi_{\sigma\omega}\, \Pi_{\rho\omega L} 
     \, \eta^2 \right .  \nonumber  \\
   & & \left. \ \ \ \ \ + (q^2 -m_\sigma^2 +\Pi_\sigma) \left[ 
     (q^2 -m_\omega^2 +\Pi_{\omega L} ) (q^2 -m_\rho^2 +\Pi_{\rho L}) 
     -\Pi_{\rho\omega L}^2  \right] \right\} 
\eeq
Mixed $\sigma$-$\delta$ meson propagation in asymmetric matter is given by
\beq
G_{\sigma\delta} &=&  {1 \over \delta_L} \left\{ 
   -(q^2 -m_\omega^2 +\Pi_{\omega L})\, \Pi_{\delta\rho}\, \Pi_{\sigma\rho}
   \, \eta^2 - (q^2 -m_\rho^2 + \Pi_{\rho L})\,  \Pi_{\delta\omega}\, 
   \Pi_{\sigma\omega}\, \eta^2  \right. \nonumber \\
   & & \ \ \ \  \ +\ (\Pi_{\delta\omega}\,
   \Pi_{\sigma\rho} + \Pi_{\delta\rho}\, \Pi_{\sigma\omega} )\, 
   \Pi_{\rho\omega L}\, \eta^2 \nonumber \\
   & & \left. \ \ \ \ \ - \left[ (q^2 -m_\omega^2 +\Pi_{\omega L} ) 
   (q^2 -m_\rho^2 +\Pi_{\rho L}) -\Pi_{\rho\omega L}^2 
    \right]\, \Pi_{\sigma\delta} \right\} 
\eeq
We have the usual $\sigma$-$\omega$ mixing in the isoscalar sector
and $\delta$-$\rho$ mixing in the isovector sectors. Morover, in
asymmetric matter there are also mixing between the isoscalar 
and isovector mesons $\delta$-$\omega$ and $\sigma$-$\rho$
\beq
G_{\sigma\omega}^\mu &=& - {\eta^\mu \over \delta_L} \left\{ 
    (\Pi_{\delta\omega} \Pi_{\sigma\rho} - \Pi_{\delta\rho} 
    \Pi_{\sigma\omega})\, \Pi_{\delta\rho}\,  \eta^2  
    -(q^2-m_\rho^2+ \Pi_{\rho L}) (q^2 -m_\delta^2 +\Pi_\delta)\, 
    \Pi_{\sigma\omega} \right. \nonumber \\
  & & \left. \ \ \ + (q^2 -m_\rho^2 +\Pi_{\rho L})\, \Pi_{\delta\omega} 
     \Pi_{\sigma\delta}\,  + (q^2 -m_\delta^2 +\Pi_\delta)\, \Pi_{\sigma\rho} 
    \, \Pi_{\rho\omega L}
     - \Pi_{\delta\rho}\, \Pi_{\sigma\delta}\, \Pi_{\rho\omega L} \right\} \\
G_{\delta\rho}^\mu &=& -{\eta^\mu \over \delta_L} \left\{ 
      ( \Pi_{\delta\omega} \Pi_{\sigma\rho} - \Pi_{\delta \rho} 
      \Pi_{\sigma \omega} )\, \Pi_{\sigma\omega} \ \eta^2 
      -(q^2 -m_\omega^2 +\Pi_{\omega L})(q^2 -m_\sigma^2 +\Pi_\sigma)
      \, \Pi_{\delta\rho} \right. \nonumber \\
   & & \left. \ \ + (q^2 -m_\sigma^2 +\Pi_\sigma)\, \Pi_{\delta\omega}\, 
    \Pi_{\rho\omega L} + (q^2 -m_\omega^2 +\Pi_{\omega L})\, 
    \Pi_{\sigma\delta}\, \Pi_{\sigma\rho} - \Pi_{\sigma\delta}\, 
    \Pi_{\sigma\omega}\, \Pi_{\rho\omega L} \right\} \\
G_{\sigma\rho}^\mu &=& -{\eta^\mu \over \delta_L} \left\{ 
    -(\Pi_{\delta\omega} \Pi_{\sigma\rho} - \Pi_{\delta\rho} 
    \Pi_{\sigma\omega})\, \Pi_{\delta\omega}\,  \eta^2 
    -(q^2 -m_\omega^2 +\Pi_{\omega L})(q^2 -m_\delta^2 +\Pi_\delta)\, 
    \Pi_{\sigma\rho} \right. \nonumber \\
    & &  \left. \ \ + (q^2 -m_\omega^2 +\Pi_{\omega L})\, \Pi_{\delta\rho} 
     \, \Pi_{\sigma\delta} + (q^2 -m_\delta^2 + \Pi_\delta )\, 
     \Pi_{\sigma\omega}\,  \Pi_{\rho\omega L}  -\Pi_{\delta\omega}\, 
     \Pi_{\sigma \delta}\, \Pi_{\rho \omega L} \right\}  \\
G_{\delta\omega}^\mu &=& -{\eta^\mu \over \delta_L} \left\{ 
     - (\Pi_{\delta\omega} \Pi_{\sigma\rho}  - \Pi_{\delta \rho} 
     \Pi_{\sigma\omega} )\, \Pi_{\sigma\rho}\,  \eta^2 
     -(q^2 -m_\rho^2 +\Pi_{\rho L}) (q^2 -m_\sigma^2 +\Pi_\sigma) 
     \, \Pi_{\delta\omega} \right. \nonumber \\
     & & \left. \ \ +(q^2 -m_\sigma^2 +\Pi_\sigma)\, \Pi_{\delta\rho}\,
      \Pi_{\rho\omega L} + (q^2 -m_\rho^2 + \Pi_{\rho L})\, \Pi_{\sigma\delta}
      \, \Pi_{\sigma\omega} - \Pi_{\sigma\rho}\, \Pi_{\sigma\delta}\, 
      \Pi_{\rho\omega L } \right\} 
\eeq
The propagators of the vector mesons $\omega$ and $\rho$ can be decomposed 
as 
\beq
G_{\omega}^{\mu\nu} &=& G_{\omega L}\, L^{\mu\nu} +  G_{\omega T}\, T^{\mu\nu}
   +  G_{\omega Q}\, Q^{\mu\nu} \nonumber \\
G_{\rho}^{\mu\nu} &=& G_{\rho L}\, L^{\mu\nu} -  G_{\rho T}\, T^{\mu\nu}
   +  G_{\rho Q}\, Q^{\mu\nu} \\
G_{\rho\omega}^{\mu\nu} &=& G_{\rho\omega L}\, L^{\mu\nu} +  G_{\rho\omega T} 
    \, T^{\mu\nu}  \nonumber 
\eeq
with for the $\omega$ meson
\beq
G_{\omega L} &=& {-1 \over \delta_L} \left\{
      (q^2 - m_\delta^2 + \Pi_\delta)\, \Pi_{\sigma\rho}^2\, \eta^2 
     + (q^2 -m_\sigma^2 + \Pi_\sigma)\, \Pi_{\delta\rho}^2\, \eta^2 
     -2\, \Pi_{\delta\rho}\, \Pi_{\sigma\rho}\, \Pi_{\sigma\delta}\, \eta^2
     \right. \nonumber \\
     & & \left. \ \ 
     +(q^2 - m_\rho^2 + \Pi_{\rho L}) \left[ (q^2 - m_\sigma^2 + \Pi_\sigma)
     (q^2 - m_\delta^2 + \Pi_\delta) - \Pi_{\sigma\delta}^2 \right]
     \right\} \\
G_{\omega T} &=& -{1 \over \delta_T} (q^2 - m_\rho^2 + \Pi_{\rho T}) \\
G_{\omega Q} &=& {1 \over m_\omega^2} 
\eeq
for the rho meson
\beq
G_{\rho L} &=& {-1 \over \delta_L} \left\{
      (q^2 - m_\delta^2 + \Pi_\delta)\, \Pi_{\sigma\omega}^2\, \eta^2 
     + (q^2 -m_\sigma^2 + \Pi_\sigma)\, \Pi_{\delta\omega}^2\, \eta^2 
     -2\, \Pi_{\delta\omega}\, \Pi_{\sigma\omega}\, \Pi_{\sigma\delta}\, 
    \eta^2 \right. \nonumber \\
     & & \left. \ \  +(q^2 - m_\omega^2 + \Pi_{\omega L}) \left[ 
     (q^2 - m_\sigma^2 + \Pi_\sigma)
     (q^2 - m_\delta^2 + \Pi_\delta) - \Pi_{\sigma\delta}^2 \right]
     \right\} \\
G_{\rho T} &=& -{1 \over \delta_T} (q^2 - m_\omega^2 + \Pi_{\omega T}) \\
G_{\rho Q} &=& {1 \over m_\rho^2} 
\eeq
and finally for the mixing between the $\omega$ and $\rho$ mesons
\beq
G_{\rho\omega L} &=& {-1 \over \delta_L} \left\{ 
     - (q^2 -m_\sigma^2 +\Pi_\sigma)\, \Pi_{\delta\rho}\, \Pi_{\delta\omega}
     \, \eta^2
     - (q^2 -m_\delta^2 + \Pi_\delta)\, \Pi_{\sigma\rho}\, \Pi_{\sigma\omega} 
      \, \eta^2 \right. \nonumber \\
   & & \ \ \ \ + ( \Pi_{\delta \omega}\, 
       \Pi_{\sigma \rho} + \Pi_{\delta \rho}\, \Pi_{\sigma\omega} )\,
      \Pi_{\sigma\delta}\,  \eta^2 \nonumber \\
   & & \left. \ \ \ \   - \left[ (q^2 -m_\sigma^2 + \Pi_\sigma)
      (q^2 -m_\delta^2 + \Pi_\delta) - \Pi_{\sigma\delta}^2 \right]
      \, \Pi_{\rho\omega L}  \right\} \\
G_{\rho\omega T} &=& - {\Pi_{\rho\omega T} \over \delta_T} 
\eeq

%%%%%%%%%%%%%%%%%%%%%%%%   5 - Numerical results    %%%%%%%%%%%%%%%%%%%%%%%%%%

\section{Numerical results}

\subsection{Choice of parameters}

At the mean field level, the equation of state is determined
by equations (\ref{eosJ},\ref{eosT}). 
The coupling constants were adjusted in order to reproduce
the properties of the saturation point. It is enough to adjust
five parameters in order to reproduce the five basic experimentally
measured properties: saturation density $n_{\rm sat}$, binding energy 
$B/A$, incompressibility modulus $K$, effective nucleon mass $m^*$ and 
asymmetry energy $a_A$. Usually these parameters are taken to be
the meson-nucleon coupling constants $g_\sigma$, $g_\omega$, 
$g_\rho$ and the non linear sigma couplings $b$, $c$; and the
coupling of the $\delta$ meson is set equal to zero. The meson masses
are kept equal to their value as quoted in the Particle Data Book
($m_\omega=782.6$ MeV, $m_\rho=769$ MeV, $m_\delta$=983 MeV),
and the $\sigma$ meson mass is taken to be equal to the standard result
of the Bonn potential model $m_\sigma=550$ MeV. The tensor coupling
of the rho meson is derivative and does not enter in the equation 
of state at the mean field level. It will be taken equal to
the Bonn potential value $f_\rho/g_\rho =6.1$ or to the vector dominance
model value $f_\rho/g_\rho = 3.7$. 

If the $\delta$ meson coupling is not equal to
zero, we have one more parameter to play with. When keeping the
parameters $g_\sigma$, $g_\omega$, $b$, $c$ constant, varying
$g_\delta$ rescales the asymmetry energy, which has to be
readjusted by changing the value of $g_\rho$. The $\delta$ is
usually discarded, moreover some studies concluded that is does 
not noticeably affect the description of stable nuclei \cite{reinhard}.
On the other hand, interest in the $\delta$ meson has been revived since
it appears that density dependent field theory needs a $\delta$ with 
a large value of the coupling in order to reproduce Dirac-Brueckner 
calculations of asymmetric matter \cite{dJL98,SST97}. Whether
or not a $\delta$ meson is present can be important for the equation 
of state of neutron stars \cite{Kutschera}. Among other things, 
a nonzero coupling to the $\delta$ manifests itself by a splitting 
of the proton and neutron effective masses. Hopefully, experiments 
planned with neutron rich nuclei will help clarifying this issue.

In principle, renormalization of vacuum fluctuations appearing in the
expressions of the nucleon effective mass and energy density should be
performed, yielding the relativistic Hartree approximation.
The procedure is standard and does not present difficulties 
(see {\it e.g.} \cite{JDAetal}). However it is known 
\cite{SW86,G89} that, after effecting a readjustment of the coupling 
constants, the equation of state is very 
similar in the renormalized (Hartree) and unrenormalized (mean field) 
case. Since the thermodynamics enters the polarizations only through 
the effective masses $M_n$, $M_p$ and chemical potentials $\mu_n$, 
$\mu_p$, we take the pragmatic point of view that it will not
change the dispersion relations whether the underlying thermodynamics
were generated by the mean field or Hartree approximations, and limit
ourselves to fit the mean field. Reasonable values for the experimental
data are 
\beq
n_{\rm sat} = 0.17\ {\rm fm}^{-3}, \quad B/A=-16\ {\rm MeV}, \quad
K=250\ {\rm MeV}, \quad m^*/m=0.8, \quad a_A=30\ {\rm MeV}
\eeq
and can be reproduced with the following parameter set
\beq
&& g_\sigma=8.00, \quad g_\omega=7.667, \quad b/g_\sigma^3=9.637\ 10^{-3}, 
\quad c/g_\sigma^4=7.847\ 10^{-3} \nonumber \\ 
&& (g_\rho,g_\delta) = (3.685, 0.)\  {\rm or}\ (5.203,5.)
\label{fiteos}
\eeq
Note that the parameter $c$ is positive, so that we are safe from
any unpleasant instabilities which may appear \cite{Wald88} 
in commonly used parametrizations where this is not the case.

This parameter set will be referred to in the following as set (1).
For the dispersion relations we will use moreover other parameter 
sets besides this one. As a matter of fact, at this level of approximation, 
there is no compelling reason why the couplings describing the thermodynamics
should be the same as those entering the RPA. When the dressed meson
propagator is used in calculations of the RPA corrections to neutrino 
opacities, the scattering processes are usually described by
parametrizations such as given in Bonn potential model. For example, 
from \cite{M87}
\beq
&& g_\sigma = 10.20, \quad g_\omega= 15.85, \quad b=0, \quad c=0 
\nonumber \\
&& g_\rho=3.19, \quad f_\rho/g_\rho=6.1, \quad g_\delta = 3.73 \\
&& \Lambda_\sigma=\Lambda_\delta=\Lambda_\rho=2000\ {\rm MeV}, 
\quad n_\rho=2, \quad \Lambda_\omega=1500\ {\rm MeV} \nonumber
\eeq
The cutoffs $\Lambda_\sigma$, $\Lambda_\omega$, $\Lambda_\rho$
and $\Lambda_\delta$ enter the definition of form factors 
which multiply the coupling constants in the calculation of the 
polarizations. They are introduced in order to take into account 
the effects of the finite size of the nucleon.
\beq
g_\alpha \rightarrow g_\alpha\, \left( {\Lambda_\alpha^2 - m_\alpha^2 \over
 \Lambda_\alpha^2 - q^2}\right)^{n_\alpha} 
\ , \quad \alpha \in \{ \sigma,\omega, \rho,\delta \} \ , \quad
n_\alpha=1\ \mbox{\rm except for}\ n_\rho=1\ {\rm or}\ 2.  \nonumber
\eeq 
This parameter set will be referred to in the following as set (2).

From \cite{BM90} we can take the parameter set of Brockmann-Machleidt 
potential C which also fits the free nucleon-nucleon scattering data, 
but with a higher value of the coupling to the $\delta$ meson
\beq
&& g_\sigma = 10.044, \quad g_\omega= 15.853, \quad b=0, \quad c=0 
\nonumber \\
&& g_\rho=3.455, \quad f_\rho/g_\rho=6.1, \quad g_\delta = 7.985 \\
&& \Lambda_\sigma=1800\ {\rm MeV}, \quad  \Lambda_\omega=\Lambda_\delta
=1500 \ {\rm MeV}, \quad \Lambda_\rho=1300\ {\rm MeV}, 
\quad n_\rho=1\nonumber
\eeq
This parameter set will be referred to in the following as set (3).

A way out of the dilemna between chosing parameters which fit the
saturation point, or the scattering data in vacuum, could be 
working with one of the recent parametrizations of the results of
Dirac-Brueckner calculations (which obtain the saturation point
using the free potential as input) in terms of density-dependent
coupling constants. A promising model is the density dependent 
relativistic hadron field theory (DDRH) which is lately
beeing developed by Lenske and collaborators \cite{FLW95}.
A serious study would involve performing linear response analysis 
on the field equations obtained from the DDRH lagrangian density.
This is currently under consideration. 

\subsection{Transverse modes}

Depending on the values of the coupling constants and cutoffs
on the one hand, and of the renormalization scheme on the other
hand, we may have a different structure of the branches. 
If the renormalization is not performed, we have spurious 
branches, as is known from previous studies \cite{JDAetal,MGP98}.
In particular, when we define the effective meson masses as 
the solution $\omega$ of the dispersion relation  on the axis
$\vec k =0$, a kink structure is found, whatever strong form 
factor is applied. When the renormalization is performed, the kink
disappears and normal branches emerge.

In asymmetric matter, the transverse mode of the omega meson mixes 
with that of the rho
meson. With renormalization scheme A (using the method of Kurasawa-Suzuki),
we obtain no more branches than the normal ones with parameters sets 
(2) or (3). On the other hand, with parameter set (1), there is also 
a zero sound branch at finite momentum, similar to the one conducing to
the (spurious) pion condensation. This zero-sound mode is related to 
the high value of the rho meson coupling constants necessary 
in order to reproduce the correct asymmetry energy, together 
with the high value of $f_\rho/g_\rho=6.1$ of the Bonn potential. 
As a matter of fact, the zero-sound mode is appreciably reduced 
when we take the vector dominance model value $f_\rho/g_\rho=3.7$. 
It disappears completely for $f_\rho=0$. On the other hand, recent
calculations from QCD sum rules \cite{Zhu} would also favor a high 
value of $f_\rho$ ($f_\rho/g_\rho = 8.0 \pm 2.0$).
The zero-sound mode is also present with parameter set (1) when
using the other renormalization scheme (B), although much weaker.
When present, the zero sound mode is somewhat reduced at larger 
proton-neutron asymmetry.
In any case, it will be possible to eliminate this spacelike mode
with a contact interaction \cite{M01} of the Landau-Migdal type (as
in the $\pi$-$\rho$ +$g'$ model), which also helps to correct 
the singular $\delta(r)$ behavior of the rho exchange potential 
at short distance.

The expression of the omega polarization is identical in schemes 
A and B. On the other hand, the result for the rho meson is quite 
different depending on the  chosen renormalization scheme.
As shown is \cite{MGP98}, the effective $\rho$ mass increases 
with increasing density in the renormalization scheme of 
Kurasawa-Suzuki (A), whereas it decreases in renormalization
scheme (B) which keeps the formal strucutre of the vacuum
terms as a function of the effective nucleon masses $M_n$, 
$M_p$.

The position of the normal branches changes with asymmetry. It is 
best seen by plotting the variation of the effective rho and omega 
meson masses with asymmetry. The effective meson masses are defined as the
solutions $\omega$ of the dispersion relation $\delta_T(\omega, 
\vec k)=0$ at vanishing three momentum $\vec k$. They
are shown on Figs. 2 and 3, (rightmost upper and lower panels), at 
four times saturation density and vanishing temperature,
with parameter set (1). 
The continuous lines were obtained by keeping
the $\omega$-$\rho$ mixing while the dashed line is obtained
by setting it to zero.

In renormalization scheme A, the omega meson effective mass
is the lowest of the two and decreases with increasing asymmetry
($\equiv$ smaller $Y_p$ values), whereas the rho meson mass increases 
with increasing asymmetry. In renormalization scheme B, the reverse 
occurs. The effective rho mass is now the lowest and decreases 
with increasing asymmetry. As in the case of the behavior of the 
$\rho$ mass with density \cite{MGP98}, the uncertainty in the choice 
of a renormalization procedure once more precludes a reliable 
prediction. 

In order to evaluate the strength of the mixing, we can define the 
mixing angle as in \cite{DMDRK97}
\beq
\tan (2\, \theta_{\omega\rho T}) = {2\, \sqrt{ | \eta^2 | }\, 
\Pi_{\rho\omega T} \over m_\rho^2 - m_\omega^2 -\Pi_{\rho T} 
+ \Pi_{\omega T} }
\eeq
The dependence of $\theta_{\omega\rho T}$ in the density and asymmetry 
parameters is represented on Fig. 4.  The curves are labelled by the value of
the proton fraction $Y_p$. The mixing angle is calculated at a fixed 
three-momentum $k=500$ MeV, at the values of $\omega$ which are 
solutions of the 
dispersion relation. We have two angles, one for each meson branch.
The mixing angle $\theta_{\omega\rho T}$ vanishes in symmetric matter
and/or at zero momentum. Non negligible values are obtained
in asymmetric matter, especially if renormalization scheme A
is used.

\subsection{Longitudinal modes}

When studying the solutions of the longitudinal dispersion relation 
$\delta_L$, we obtain the normal branches corresponding to each of 
the four mesons involved in the longitudinal part of the dispersion 
relation. Again, the behavior of the corresponding effective masses 
as a function of density depends on the choice of the coupling 
constants, cutoffs and renormalization scheme. In both schemes, 
the sigma meson mass increases with density, more rapidly in scheme B. 
The rho meson mass increases in scheme A and decreases in scheme B. 
The delta meson mass decreases in scheme A and increases in scheme B. 
The omega meson mass coincides in both schemes, it first decreases 
with density and then starts increasing.

The behavior of the effective masses at a fixed density 
$n_B=4\ n_{\rm sat}$ as a function of the proton fraction $Y_p$ is
represented in Figs. 2 and 3. The effective masses obtained for the 
omega and rho meson from the longitudinal dispersion relation coincide 
with those obtained from the transverse modes, as it should be.
The delta mass is larger at smaller $Y_p$ in both schemes. The
sigma mass increases in scheme B but first decreases and then slightly 
increases with asymmetry in scheme A.

Besides the normal branches, we have also heavy rho meson branches
(see also {\it e.g.} \cite{MGP98})
which would disappear if a stronger cutoff were applied. These will 
not be further studied here. Finally, there are zero sound branches.
The zero-sound branch due to $\sigma$-$\omega$ mixing is already
well known from studies of symmetric matter. It is present in both 
renormalization schemes with parameter sets (2) and (3). 

In Fig. 5 we show the dependence of the zero-sound mode due to 
$\sigma$-$\omega$ mixing with the asymmetry, at four times saturation 
density and vanishing temperature. It is seen that at high asymmetry
the strength of the zero sound is only slightly reduced. The associated
characteristic velocity increases, resulting in a small rotation
of the zero sound branch in the $\omega$-$k$ plane. 
When studying the imaginary parts of the polarization, it is seen that 
at zero temperature the lower branch of the zero sound mode is 
suppressed by Landau damping, whereas the higher branch is not. 

In the delta-rho sector, no zero sound branch similar to that existing 
in the $\sigma$-$\omega$ system was found, whatever parameter set was used. 
Some spurious branches may appear in the spacelike region at high momentum, 
if we work with parameter set (1) and in renormalization scheme A. 
They disappear when applying a moderate cutoff.  

\subsection{Effect of the mixing on the propagators}

In the preceding paragraphs, we obtained results for the effective 
meson masses and mixing angles, which concern the timelike region 
of the $\omega$-$k$ plane. In order to characterize the influence 
of meson mixing in the spacelike part of the $\omega$-$k$ plane,
we also studied the propagators at fixed $k$ for
varing $\omega$ with the condition $\omega<k$. The knowledge of 
the  behavior of the propagators for this parameter range can be 
of use for the study of RPA corrections to neutrino-nucleon 
scattering, since it enters the definition of the RPA correction
to the polarization insertion of Eq. (\ref{Dyson}).

A sample of the results is displayed in Figs. 6 to 8. The thermodynamical
conditions were chosen to be $n_B=4\, n_{\rm sat}$, T=20 MeV, neutrino
free matter in $\beta$ equilibrium $Y_\nu=0$. The exchanged
3-momentum was fixed at $k=50$ MeV. Similar features were observed 
for other values of $n_B$ and $k$. 

The real and imaginary part of the $\sigma$ meson propagator are shown 
on Fig. 6. The $G_{\sigma\omega}$ and $G_{\omega\, L}$ behave in the 
same way. The behavior of the real part of $G_{\sigma}$ is dominated
by the pole at the location of the zero sound branch. The lower part 
of the zero sound  branch falls into the region of Landau damping 
($\omega/k=0.67$ on the figure) so that it does not give rise to a pole.
On the other hand, the upper branch falls on the fringe of the Landau 
damping zone at zero temperature, so that the corresponding pole
($\omega/k=0.83$ on the figure) remains largely unscreened. 
The effect of mixing is not very strong.

As we saw in the preceding paragraph, there does not exist a
$\delta$-$\rho$ zero sound branch. Due to the full 
$\sigma$-$\omega$-$\rho$-$\delta$ mixing however, the propagators
of the $\delta$-$\rho$ subsystem get contaminated by the
zero sound pole in the $\sigma$-$\omega$ subsystem, as can be seen
on Fig. 7, where we plotted the $G_{\delta\rho}$ propagator.
The $G_{\delta}$ and  $G_{\rho\, L}$ propagators behave in a 
similar way.

Finally, Fig. 8 shows an example of the behavior of the transverse rho 
propagator $G_{\rho\, T}$. The mixing with the $\omega$ meson
is responsible for a moderate modification of the real and imaginary 
parts.

\section{Conclusion}

The dispersion relations of the neutral mesons $\sigma$, $\omega$, $\rho$
and $\delta$ have been obtained in the random phase approximation
in the framework of a relativistic hadronic field theory. 
It was shown that in nuclear matter with a proton-neutron asymmetry, 
new mixing channels open between mesons of different isospin number. 
A numerical evaluation indicates that the influence of this mixing 
on the effective meson masses and zero sound mode is moderate, 
of the order of 
5 \% to 10 \%. An application of these to the calculation 
of RPA corrections to neutrino-nucleon scattering also leads to effects 
of this order of magnitude. A similar study concerning the charged mesons
is presently underway.

The amount of corrections arising from meson mixing is somewhat 
larger in  the $\rho$-$\omega$ (both transverse and longitudinal) and
$\delta$-$\omega$ channels than in the $\rho$-$\sigma$ or 
$\sigma$-$\delta$ channels for the thermodynamical and kinematic 
conditions explored in this work. The role of these new mixing 
channels on the behavior of the dilepton production rate in 
heavy ion collisions \cite{DMDRK97,BF98}  remains to be investigated, 
and could bring further arguments to the discussion of the behavior of
the $\rho$ meson in the medium.

\vskip 1cm

\noindent{\Large{\bf Acknowledgement}}

\vskip 0.5cm

This work was supported in part by the spanish grant 
n$^{\underline{\rm o}}$ MCT-00-BFM-0357.

\newpage

\noindent{\Large{\bf Appendix A} - Real, Matter Part of Polarizations}

\vskip 0.5cm

The general expressions of the matter part of the  polarizations can be 
written in terms of the integrals for $i \in \{ p,n \}$
\beq
I_{m\, (i)}(p,\omega,k) &=& \int_{-1}^{1} du {(p u )^n \over 
( \omega \sqrt{p^2 + M_i^2} - p k u -q^2/2)(\omega \sqrt{p^2 + M_i^2} - p k u 
+q^2/2)} \\
{\cal I}^n_{m\, (i)}(\omega,k) &=& \int_0^\infty \ {p^2 dp  \over
\sqrt{p^2 + M_i^2}}\ 
(p^2+M_i^2)^n\, I_m(p,\omega,k)\ \Bigl(n_i(p) + (-1)^{n+m} 
\overline n_i(p) \Bigr)   \\
X_0^{(i)} &=& \int_0^\infty\ {p^2 dp \over \sqrt{p^2 + M_i^2}}
\ \Bigl( n_i(p) + \overline n_i(p) \Bigr)
\eeq
Expressions for $\Pi_{\sigma}$, $\Pi_{\omega}^{\mu\nu}$,  
$\Pi_{\sigma\omega}^{\mu}$,  $\Pi_\delta$, $\Pi_\rho^{\mu\nu}$ and
$\Pi_{\delta\rho}^\mu$ have already been given in Appendix A of \cite{M01} 
in the case of symmetric nuclear matter. In asymmetric matter, one should
remove the degeneracy factor $d=2$  of the formulae presented in
\cite{M01} and sum instead on contributions of the proton and neutron loops.
For example, the $\delta$-$\rho$ mixing polarization is now given by
\beq
{\cal R}e\ \Pi_{\delta\rho}&=& {\cal R}e \Pi_{\delta\rho\, (p)} + 
{\cal R}e\ \Pi_{\delta\rho\, (n)}\nonumber \\
{\cal R}e\ \Pi_{\delta\rho\, (i)}^\mu&=&-{g_\delta\over 4 \pi^2} 
(2 M_i\ g_\rho + {f_\rho \over 2 m} q^2) q^2 \eta^\mu 
\left[ {\cal I}^1_{0\, (i)} -{\cal I }^0_{1\, (i)} \right] 
\eeq
In asymmetric matter, we will moreover have mixing polarizations between 
mesons of different isospin number. They are given by the difference
between the contributions of the proton and neutron loops.
\beq
{\cal R}e\ \Pi_{\omega\rho L}&=& {\cal R}e\  \Pi_{\omega\rho L\, (p)}
- {\cal R}e\ \Pi_{\omega\rho L\, (n)} \nonumber\\
{\cal R}e\ \Pi_{\omega\rho L\, (i)}&=& - {g_\omega g_\rho \over 2 \pi^2} 
\left[ {\cal I}^2_{0\, (i)}  -{\cal I }^0_{2\, (i)} \right] 
-\left( {f_\rho \over 2 m} \right) {g_\omega \over 4 \pi^2} M_i\
q^4 {\cal I }^0_{0\, (i)} \nonumber \\
{\cal R}e\ \Pi_{\omega\rho T} &=&   {\cal R}e\  \Pi_{\omega\rho T\, (p)}
- {\cal R}e\ \Pi_{\omega\rho T\, (n)} \nonumber\\
{\cal R}e\ \Pi_{\omega\rho T\, (i)} &=&  - {g_\omega g_\rho \over 4 \pi^2} 
\left[M_i^2\, q^2 {\cal I }^0_{0\, (i)} + (\omega^2 + k^2) 
{\cal I}^2_{0\, (i)}  +{\cal I }^0_{2\, (i)} -4 \omega k\, 
{\cal I }^1_{1\, (i)} \right] \nonumber \\
{\cal R}e\ \Pi_{\delta\omega}^\mu &=&  {\cal R}e\  \Pi_{\delta\omega (p)}^\mu
- {\cal R}e\ \Pi_{\delta\omega (n)}^\mu \nonumber\\
{\cal R}e\ \Pi_{\delta\omega\, (i)}^\mu &=&- {g_\delta g_\omega \over 2 \pi^2}
\eta^\mu M_i\ q^2  \left[ {\cal I}^1_{0\, (i)} -{\omega \over k}
 {\cal I}^0_{1\, (i)} \right] \nonumber \\
{\cal R}e\ \Pi_{\sigma\rho}^\mu &=& {\cal R}e\  \Pi_{\sigma\rho (p)}^\mu
- {\cal R}e\ \Pi_{\sigma\rho (n)}^\mu \nonumber\\
{\cal R}e\ \Pi_{\sigma\rho\, (i)}^\mu &=& -g_\sigma \left( 2 M_i\ g_\rho 
+ {f_\rho \over 2 m} q^2 \right){q^2 \over 4 \pi^2} 
\left[ {\cal I}^1_{0\, (i)} -{\omega \over k}
 {\cal I}^0_{1\, (i)} \right] \nonumber \\
{\cal R}e\ \Pi_{\sigma\delta} &=& {\cal R}e\  \Pi_{\sigma\delta (p)}
- {\cal R}e\ \Pi_{\sigma\delta (n)} \nonumber\\
{\cal R}e\ \Pi_{\sigma\delta\, (i)} &=& - {g_\sigma g_\delta \over 2 \pi^2}
\left[ 2 X_0^{(i)} +\left( {q^4 \over 4} -M_i^2\, q^2  {\cal I }^0_{0\, (i)} 
\right) \right] \nonumber
\eeq 

\newpage

\noindent{\Large{\bf Appendix B} - Renormalization}

\vskip 0.5cm

Let us first consider the scalar mesons:

The contribution of the vacuum to the $\sigma$ meson polarization is
\beq
{\cal R}e\ \Pi_{\sigma\sigma}^{\rm vac} &=& {g_\sigma^2 \over 2 \pi^3}
\left[ {\cal I}_1^{(p)} + \left( {q^4 \over 4} -M_p^2 q^2 \right)
{\cal I}_2^{(p)} + {\cal I}_1^{(n)} + \left( {q^4 \over 4} -M_n^2 q^2 \right)
{\cal I}_2^{(n)} \right] \nonumber \\ 
{\rm with}\quad
{\cal I}_1^{(i)} & = & \int d^4 p\ \delta(p^2-M_i^2)\, \theta(-p_0) 
\nonumber \\ 
{\cal I}_2^{(i)} & = &  \int d^4 p\ {\delta(p^2 -M_i^2)\, \theta(-p_0) 
\over (p.k)^2 -k^4/4} \nonumber
\eeq
The vacuum part of the $\delta$ polarization 
${\cal R}e \Pi_{\delta\delta}^{\rm vac}$ has the same structure.
The divergent part of the integrals ${\cal I}_1^{(i)}$ and ${\cal I}_2^{(n)}$
is extracted by dimensional regularization
\beq
{\cal I}_1^{(i)\ {\rm reg}} &=& \pi M_i^2 \left[ -{1 \over \epsilon} + \ln 
\left( {M_i \over m} \Lambda_1 \right) \right] \nonumber \\
{\cal I}_2^{(i)\ {\rm reg}} &=& -{2 \pi \over q^2} \left[ -{1 \over \epsilon} 
+ \ln \left( {M_i \over m} \Lambda_2 \right) -{1 \over 2} + \theta(q^2, M_i^2)
\right] \nonumber 
\eeq
where $\theta$ is a finite valued function
\beq
\theta(q^2, M_i^2) = y \int_0^\infty {dx \over (x^2+y)\sqrt{x^2+1}}
\qquad {\rm with}\ y=1-{q^2 \over 4 M_i^2} 
\eeq
In order to simplify somewhat the expressions, we will use the following 
short notations 
\beq
\theta_p &=& \theta(q^2,M_p^2) \quad , \ \theta_n = \theta(q^2,M_n^2)
\nonumber \\
\theta_\sigma &=& \theta(m_\sigma^2,m^2) \quad ,\ \theta_\delta= 
\theta(m_\delta^2,m^2) \nonumber \\
\theta_{\sigma q} &=& {\partial \theta(q^2,M_i^2) \over \partial q^2}
|_{q^2=m_\sigma^2, M_i^2=m^2}
\label{thetafon}
\eeq
and work in units of the free nucleon mass ({\it i.e.}, put $m=1$).
$\epsilon$ is a vanishingly small quantity and $\Lambda_1$, $\Lambda_2$
are finite arbitrary integration constants.

We obtain infinities of the type
\beq
\Pi_{\sigma\sigma}^{\rm vac}={g_\sigma^2 \over 2 \pi^2} \left[ {M_p^2 +M_n^2 
\over \epsilon_1} + {q^2 \over \epsilon_2} + \Lambda_{\sigma\sigma}
\right] \nonumber
\eeq
with $M_n^2 + M_p^2= 2 ( m -g_\sigma \sigma)^2 + 2 g_\delta^2 \delta^2 $
and $\Lambda_{\sigma\sigma}$ is a finite contribution.
Since the mixed polarization $\Pi_{\sigma\delta}^{\rm vac}$ depends on 
the difference of the contributions of proton and neutron loops, one finds 
that the $q^2/\epsilon$ singularities from both contributions cancel 
each other, and one is left with
\beq
\Pi_{\sigma\delta}^{\rm vac}={g_\sigma g_\delta \over 2 \pi^2} \left[ 
{M_p^2 -M_n^2 \over \epsilon_3} + \Lambda_{\sigma\delta}
\right] \nonumber
\eeq
with $M_p^2 -M_n^2 = 2 (m-g_\sigma \sigma) g_\delta \delta$
and $\Lambda_{\sigma\delta}$ is a finite contribution.
$\Pi_{\delta\delta}^{\rm vac}$ and $\Pi_{\delta\sigma}^{\rm vac}$
have similar structures.

These infinite contributions can be cancelled by adding to the Lagrangian
(\ref{Lagrangian}) the counterterms
\beq
Z_\sigma \partial_\mu \sigma \partial^\mu \sigma  +
Z_\delta \partial_\mu \delta \partial^\mu \delta 
+ {1 \over 2} A_1 \sigma^2 + {1 \over 3} A_2 \sigma^3 
+ {1 \over 4} A_3 \sigma^4 + B_1 \delta_2 + B_2 \sigma \delta^2 
+ B_3 \sigma^2 \delta^2 +C \delta^4
\eeq
which contribute to the dispersion relation as {\it e.g.}
\beq
(-q^2 + M_\sigma^2) \sigma_1 &=&\left[ \Pi_{\sigma\sigma}^{\rm mat} 
+ {g_\sigma^2 \over 2 \pi^2} \Lambda_{\sigma\sigma} \right] \sigma_1 
+ \left[ \Pi_{\sigma\delta}^{\rm mat}  + {g_\sigma g_\delta \over 2 \pi^2} 
\Lambda_{\sigma\delta} \right] \delta_1
\nonumber \\
&& + \left[ Z_\sigma q^2 + A_1 + 2
A_2 \sigma + 3 A_3 \sigma^2 + 2 B_3 \delta^2 +{ g_\sigma^2 \over 2
\pi^2} \left\{ { 2 (m -g_\sigma \sigma)^2 + 2 g_\delta^2 \delta^2 
\over \epsilon_1} + {q^2 \over \epsilon_2}
\right\} \right] \sigma_1 \nonumber \\
&& +\left[ 2 B_2 \delta + 4 B_3 \sigma \delta + {g_\sigma g_\delta 
\over 2 \pi^2}
\left\{ {2 (m -g_\sigma \sigma) g_\delta \delta \over \epsilon_3} \right\} 
\right] \delta_1 \nonumber
\eeq
The infinities are neutralized by choosing the couplings of the
counterterm Lagrangian so that, for example
\beq
& 2 B_3 + \displaystyle{g_\sigma^2 g_\delta^2 \over \pi^2} 
\displaystyle{1 \over \epsilon_1 } 
= b_3^{\sigma\sigma} &  \nonumber \\
& 4 B_3 - \displaystyle{g_\sigma^2 g_\delta^2 \over \pi^2}
\displaystyle{1 \over \epsilon_3}
= b_3^{\sigma\delta} &  \qquad 2 B_2 +
m \displaystyle{g_\sigma g_\delta^2 \over \pi^2}
\displaystyle{1 \over \epsilon_3} = b_2^{\sigma\delta} \nonumber \\
& 2 B_3 + \displaystyle{g_\sigma^2 g_\delta^2 \over \pi^2}
\displaystyle{1 \over \epsilon_4} 
= b_3^{\delta\delta} &  \qquad  2 B_2 -2 m  \displaystyle{g_\sigma g_\delta^2 
\over \pi^2} \displaystyle {1 \over \epsilon_4} 
= b_2^{\delta\delta} \label{compat1} \\
& 4 B_3 - \displaystyle{g_\sigma^2 g_\delta^2 \over \pi^2}{
\displaystyle1 \over \epsilon_6}
= b_3^{\delta\sigma} & \qquad  2 B_2 +m \displaystyle{g_\sigma 
g_\delta^2 \over \pi^2}
\displaystyle{1 \over \epsilon_6} = b_2^{\delta\sigma} \nonumber
\eeq
with $b_3^{\sigma\sigma}$, $ b_2^{\sigma\delta}$, ... {\it etc}
finite. The unknown finite remnants are defined so that
the vacuum polarizations
\beq
\Pi_{\sigma\sigma}^{\rm vac} &=&  {g_\sigma^2 \over 2 \pi^2} 
\Lambda_{\sigma\sigma} + z_\sigma q^2 + a_1 + 2 a_2 \sigma + 3 a_3 \sigma^2 +
2 b_3^{\sigma\sigma} \delta^2 \nonumber \\
\Pi_{\sigma\delta}^{\rm vac} &=&  {g_\sigma g_\delta \over 2 \pi^2} 
\Lambda_{\sigma\delta} + 2 b_2^{\sigma\delta} \delta + 4 b_3^{\sigma\delta} 
\nonumber
\eeq
fulfill some physical conditions. One generally requires the polarization
$\Pi_{\sigma\sigma}^{\rm vac}$ to cancel in the vacuum 
($M_p=M_n=m$, $\sigma=\delta=0$) on the mass shell $q^2=m_\sigma^2$.
In \cite{JDAetal} one further imposes that the first derivatives with
respect to the $\sigma$ and $\delta$ fields vanish at the same point
\beq
{\partial \Pi_{\sigma\sigma}^{\rm vac} \over \partial_\sigma}
|_{q^2=m_\sigma^2, M_p=M_n=m, \sigma=\delta=0}=0\ , \quad
{\partial^2 \Pi_{\sigma\sigma}^{\rm vac} \over \partial_\sigma^2}
|_{q^2=m_\sigma^2, M_p=M_n=m, \sigma=\delta=0}=0
\label{renJDA}
\eeq
In the scheme of Kurasawa and Suzuki (scheme A in this work), 
one requires that these derivatives vanish at $q^2=0$
\beq
{\partial \Pi_{\sigma\sigma}^{\rm vac} \over \partial_\sigma}
|_{q^2=0, M_p=M_n=m, \sigma=\delta=0}=0\ , \quad
{\partial^2 \Pi_{\sigma\sigma}^{\rm vac} \over \partial_\sigma^2}
|_{q^2=0, M_p=M_n=m, \sigma=\delta=0}=0
\label{renKS}
\eeq
In the schemes of Shiomi and Hatsuda \cite{SH94} or scheme 3 of 
\cite{MGP98} (scheme B of this work), the structure of the
expressions in terms of $M_n$, $M_p$ is preserved, leading to 
conditions such as
\beq
a_1  + 2 a_2 \sigma + 3 a_3 \sigma^2  +
2 b_3^{\sigma\sigma} \delta^2= a_1 {1 \over 2 m^2} 
\left[ 2 (m-g_\sigma \sigma)^2 + 2 g_\delta^2 \delta^2 \right]
=  a_1 {1 \over 2 m^2} \left[M_n^2 + M_p^2 \right]
\label{renschB}
\eeq
These conditions yield expressions of the constants  
$b_3^{\sigma\sigma}$, 
$ b_2^{\sigma\delta}$, ... {\it etc} in terms of the function $\theta$
defined in Eqs. (\ref{thetafon}). For example,  in scheme B one obtains
\beq
b_2^{\sigma\delta} &=& {g _\sigma g_\delta^2 \over 2 \pi^2} m \left(
-4 \theta_\sigma + m_\sigma^2 (4-m_\sigma^2) \theta_{\sigma q}
\right) = {g _\sigma g_\delta^2 \over 2 \pi^2} m x_\sigma 
\quad , \qquad b_3^{\sigma\delta}=
- {g _\sigma^2 g_\delta^2 \over 2 \pi^2}  x_\sigma
\nonumber \\
b_2^{\delta\delta} &=& - {g _\sigma g_\delta^2 \over  \pi^2} m \left(
-4 \theta_\delta + m_\delta^2 (4-m_\delta^2) \theta_{\delta q}
\right) = - {g _\sigma g_\delta^2 \over  \pi^2} m x_\delta
\quad , \qquad b_3^{\delta\delta} = {g _\sigma^2 g_\delta^2 
\over 2 \pi^2} x_\delta \nonumber
\eeq
The constants determined by these conditions must moreover fulfill the
compatibility conditions imposed by
(\ref{compat1})
\beq
b_3^{\sigma\delta} - b_3^{\delta\sigma} ={g_\sigma \over m} \left( 
b_2^{\delta\sigma}- b_2^{\sigma\delta} \right)
\quad , \qquad 2 b_3^{\delta\delta} - b_3^{\sigma\delta}
= {g_\sigma \over m} \left( b_2^{\sigma\delta} - b_2^{\delta\delta}
\right)
\label{compat2}
\eeq
The constants $b_3^{\sigma\delta}$, $ b_3^{\delta\sigma}$, 
$b_2^{\delta\sigma}$ as determined by the sets of conditions
(\ref{renKS}) or (\ref{renschB}) do indeed fulfill these compatibility
conditions, whereas the set of conditions (\ref{renJDA}) do not.

We now give the full expressions of the renormalized polarizations in 
schemes A or B.

\vskip 0.5cm
$\underline{\mbox{\sf Renormalization scheme A}}$  
\vskip 0.5cm

\beq 
\Pi_{\sigma\sigma}^{\rm vac\ (A)} &=& {g_\sigma^2 \over 2 \pi^2} \Bigg[ 
  3\, M_p^2 \ln M_p + 3\, M_n^2 \ln M_n 
  -{q^2 \over 2} \left( \ln M_p + \theta_p \right) 
 -{q^2 \over 2} \left( \ln M_n + \theta_n \right) 
 \nonumber \\
& & \qquad +2\, M_p^2 \theta_p  
+2\, M_n^2 \theta_n + (q^2 - m_\sigma^2) \left(\theta_\sigma - 
(4 -m_\sigma^2) \theta_{\sigma q} \right) - (4 - m_\sigma^2) \theta_\sigma
\nonumber \\
& & \qquad + \left( 1 -13 {M_n^2 + M_p^2 \over 2} +6 (M_n +M_p) \right) \Bigg]
\eeq
The expression of $\Pi_{\delta\delta}^{\rm vac\ (A)}$ can be obtained by
replacing everywhere $g_\sigma$, $m_\sigma$ $\theta_\sigma$, $\theta_{\sigma q}$ 
by $g_\delta$, $m_\delta$ $\theta_\delta$, $\theta_{\delta q}$ 
in $\Pi_{\sigma\sigma}^{\rm vac\ (A)}$.
\beq
\Pi_{\sigma\delta}^{\rm vac\ (A)} &=& {g_\sigma g_\delta \over 2 \pi^2} \Bigg[ 
  3\, M_p^2 \ln M_p - 3\, M_n^2 \ln M_n -{q^2 \over 2} \left( \ln M_p 
+ \theta_p \right) +{q^2 \over 2} \left( \ln M_n + \theta_n\right) 
 \nonumber \\
& & \qquad  +2\, M_p^2 \theta_p  -2\, M_n^2 \theta_n 
+ \left( { M_n - M_p \over 2} \right) \left( - 12 + 13(M_n + M_p)
\right) \Bigg] 
\eeq
Again, the expression of $\Pi_{\delta\sigma}^{\rm vac\ (A)}$ is obtained by
replacing everywhere the indices $\sigma$ by $\delta$ in 
$\Pi_{\sigma\delta}^{\rm vac\ (A)}$.

For the vector mesons, we have
\beq
\Pi_{\omega\omega}^{{\rm vac\ (A)}\ \mu\nu} &=& 
{g_\omega^2 \over 6 \pi^2} \left[
2 M_p^2 (\theta_p -1) + 2 M_n^2 (\theta_n -1) + q^2  \left( \ln M_p + \theta_p
\right) + q^2 \left( \ln M_n + \theta_n \right) \right. \nonumber \\
& & \left. \qquad -{4 q^2 \over m_\omega^2} ( \theta_\omega -1) 
-2 q^2 \theta_\omega \right] 
\times \left( g^{\mu\nu} -{q^\mu q^\nu \over q^2} \right)
\eeq

\beq
\Pi_{\omega\rho}^{{\rm vac\ (A)}\ \mu\nu} &=& 
\Pi_{\rho\omega}^{{\rm vac\ (A)}\ \mu\nu} 
\nonumber \\
&=& \left\{ {g_\omega g_\rho \over 6 \pi^2} \left[
2 M_p^2 (\theta_p -1) - 2 M_n^2 (\theta_n -1) + q^2  \left( \ln M_p + \theta_p
\right) - q^2 \left( + \ln M_n + \theta_n \right) \right] \right. \nonumber \\
&& \left.  + \left( {f_\rho \over 2 m} \right) {g_\omega \over 2 \pi^2} 
q^2 \left[  M_p \left( \ln M_p + \theta_p \right) 
- M_n \left( \ln M_n + \theta_n \right) +2 (M_n - M_p) \right] \right\}
\nonumber \\
& & \qquad \qquad 
\times \left( g^{\mu\nu} -{q^\mu q^\nu \over q^2} \right)
\eeq

\beq
\Pi_{\rho\rho}^{{\rm vac\ (A)}\ \mu\nu} &=& 
\left\{ {g_\rho^2 \over 6 \pi^2} \Bigl[
2 M_p^2 (\theta_p -1) + 2 M_n^2 (\theta_n -1)+
 q^2  \left( \ln M_p + \theta_p \right) + q^2 \left( \ln M_n + \theta_n
 \right)  \right. \nonumber \\
& &  \qquad \quad -{4 q^2 \over m_\rho^2} ( \theta_\rho -1) 
-2 q^2 \theta_\rho \Bigr] \nonumber \\
&& + \left( {f_\rho \over 2 m} \right)^2 {q^2 \over 6 \pi^2} \left[ 3 M_p^2
  \ln M_p + 4 M_p^2 \theta_p + {q^2 \over 2}  \left( \ln M_p + \theta_p
  \right) + 3 M_n^2  \ln M_n + 4 M_n^2 \theta_n \right. \nonumber \\
&&  \qquad \qquad  + {q^2 \over 2}   \left( \ln M_n + \theta_n \right)
 -(8 +m_\rho^2) \theta_\rho -(q^2 - m_\rho^2) (\theta_\rho 
+ (8 +m_\rho^2) \theta_{\rho q} ) \nonumber \\
&& \left. \qquad \qquad +5 +6 (M_n+M_p) -17 {M_p^2 +M_n^2 \over 2}
\right] \nonumber \\
&& \left. + \left( {f_\rho \over 2 m} \right) {g_\rho \over  \pi^2} q^2 \left[
 M_p \left( \ln M_p + \theta_p \right) + M_n \left( \ln M_n + \theta_n \right)
-2 \theta_\rho +4 \left( m -{M_n + M_p \over 2} \right) \right]
 \right\} \nonumber \\
&& \qquad \qquad
\times \left( g^{\mu\nu} -{q^\mu q^\nu \over q^2} \right)
\eeq

\vskip 0.5cm
$\underline{\mbox{\sf Renormalization scheme B}}$  
\vskip 0.5cm

\beq 
\Pi_{\sigma\sigma}^{\rm vac\ (B)} &=& {g_\sigma^2 \over 2 \pi^2} \Bigg[ 
  3\, M_p^2 \ln M_p + 3\, M_n^2 \ln M_n  -{q^2 \over 2} \left( \ln M_p 
+ \theta_p\right) -{q^2 \over 2} \left( \ln M_n + \theta_n\right) 
\nonumber \\
& & \qquad +2\, M_p^2 \theta_p 
+2\, M_n^2 \theta_n  + \left( 1- { M_p^2 + M_n^2 \over 2} \right) 
\left( 4 \theta_\sigma -m_\sigma^2 (4 -m_\sigma^2) \theta_{\sigma q} 
\right)\nonumber \\
& & \qquad + (q^2 - m_\sigma^2) \left( \theta_\sigma - (4-m_\sigma^2) 
\theta_{\sigma q} \right)  -(4 -m_\sigma^2) \theta_\sigma \Bigg] 
\eeq
\beq
\Pi_{\sigma\delta}^{\rm vac\ (B)} &=& {g_\sigma g_\delta \over 2 \pi^2} \Bigg[ 
  3\, M_p^2 \ln M_p - 3\, M_n^2 \ln M_n -{q^2 \over 2} \left( \ln M_p 
+ \theta_p \right) +{q^2 \over 2} \left( \ln M_n + \theta_n \right) 
\nonumber \\
& & \qquad  +2\, M_p^2 \theta_p  -2\, M_n^2 \theta_n 
+ \left( { M_p^2 - M_n^2 \over 2} \right) \left( - 4 \theta_\sigma
+m_\sigma^2 (4 -m_\sigma^2) \theta_{\sigma q} \right) \Bigg] \\
\eeq
Once more, the expressions of $\Pi_{\delta\delta}^{\rm vac\ (B)}$ and 
$\Pi_{\delta\sigma}^{\rm vac\ (B)}$ are obtained by
replacing everywhere the indices $\sigma$ by $\delta$ in 
$\Pi_{\sigma\sigma}^{\rm vac\ (B)}$ and $\Pi_{\sigma\delta}^{\rm vac\ (B)}$
respectively.

The expression of $\Pi_{\omega\omega}^{{\rm vac}\mu\nu}$ 
coincides in both schemes.
For the polarizations involving a vertex with the rho meson, we have
\beq
\Pi_{\omega\rho}^{{\rm vac (B)}\ \mu\nu} &=& 
\left\{ {g_\omega g_\rho \over 6 \pi^2} \left[
2 M_p^2 (\theta_p -1) - 2 M_n^2 (\theta_n -1) + q^2  \left( \ln M_p + \theta_p
\right) - q^2 \left( + \ln M_n + \theta_n \right) \right] \right. \nonumber \\
&& \left.  + \left( {f_\rho \over 2 m} \right) {g_\omega \over 2 \pi^2} 
q^2 \left[  M_p \left( \ln M_p + \theta_p \right) 
- M_n \left( \ln M_n + \theta_n \right) + (M_n - M_p) \theta_\omega \right] 
\right\} \nonumber \\
& & \qquad \qquad 
\times \left( g^{\mu\nu} -{q^\mu q^\nu \over q^2} \right)
\eeq
and $\Pi_{\rho\omega}^{{\rm vac\ (B)}\ \mu\nu}$ is obtained by replacing
$\theta_\omega$ by $\theta_\rho$ in the last term.

\newpage

\beq
\Pi_{\rho\rho}^{{\rm vac\ (B)}\ \mu\nu} &=& 
\left\{ {g_\rho^2 \over 6 \pi^2} \Bigl[
2 M_p^2 (\theta_p -1) + 2 M_n^2 (\theta_n -1)+
 q^2  \left( \ln M_p + \theta_p \right) + q^2 \left( \ln M_n + \theta_n
 \right)  \right. \nonumber \\
& &  \qquad \quad -{4 q^2 \over m_\rho^2} ( \theta_\rho -1) 
-2 q^2 \theta_\rho \Bigr] \nonumber \\
&& + \left( {f_\rho \over 2 m} \right)^2 {q^2 \over 6 \pi^2} \left[ 3 M_p^2
  \ln M_p + 4 M_p^2 \theta_p + {q^2 \over 2}  \left( \ln M_p + \theta_p
  \right) + 3 M_n^2  \ln M_n + 4 M_n^2 \theta_n \right. \nonumber \\
&&  \qquad \qquad  + {q^2 \over 2}   \left( \ln M_n + \theta_n \right)
 -(8 +m_\rho^2) \theta_\rho -(q^2 - m_\rho^2) (\theta_\rho 
+ (8 +m_\rho^2) \theta_{\rho q} ) \nonumber \\
&& \left. \qquad \qquad + \left( {M_p^2 + M_n^2 \over 2} -1 \right)
\left( -8 \theta_\rho + m_\rho^2 (8 +m_\rho^2) \theta_{\rho q} \right)
\right] \nonumber \\
&& \left. + \left( {f_\rho \over 2 m} \right) {g_\rho \over  \pi^2} q^2 \left[
 M_p \left( \ln M_p + \theta_p \right) + M_n \left( \ln M_n + \theta_n \right)
- \theta_\rho (M_n + M_p) \right]
 \right\} \nonumber \\
&& \qquad \qquad
\times \left( g^{\mu\nu} -{q^\mu q^\nu \over q^2} \right)
\eeq

As mentioned before, the renormalization constants fulfill 
the compatibility conditions (\ref{compat2}).
Note however that, in scheme B, $\Pi_{\sigma\delta} \not= \Pi_{\delta\sigma}$ 
since they are renormalized on the mass shell of the $\sigma$ and $\delta$ 
respectively, {\it i.e.}
\beq
\Pi_{\sigma\delta} (q^2 = m_\sigma^2)=0 \quad ; 
\Pi_{\delta\sigma} (q^2 = m_\delta^2)=0 \quad \nonumber
\eeq
This breaks the symmetry which was explicit on the initial divergent
expression. The symmetry could be reestablished {\it e.g.} by imposing
that the polarizations vanish at some intermediate common scale 
$\mu_{\sigma\delta}$
\beq
\Pi_{\sigma\delta} (q^2 = \mu_{\sigma\delta}^2)=0= 
\Pi_{\delta\sigma} (q^2 = \mu_{\sigma\delta}^2)\nonumber
\eeq
In renormalization scheme A on the other hand the symmetry was already 
preserved by the choice of the common renormalization point $q^2=0$.
The same remark applies to $\Pi_{\omega\rho}^{\mu\nu} \not= 
\Pi_{\rho\omega}^{\mu\nu}$ in scheme B since they are renormalized 
on the mass shell of the $\omega$ and $\rho$  respectively.
They coincide in scheme A.

\vskip 1cm

\noindent{\Large{\bf Appendix C} - Imaginary Parts of Polarizations}

\vskip 0.5cm
The imaginary part arise the factor $\pm i \epsilon$ in the
denominator of (\ref{propagator}) after applying the formula
$1/(x+i \epsilon) = {\cal P}/x -i \pi \delta(x)$.
Depending on the prescription applied for going around the 
pole in the propagator, we obtain the imaginary part
of one of the definitions $\overline\Pi$, $\Pi^R$, $\Pi^<$,
$\Pi^{11}$ ...  all these being related to each other through factors 
such  as $\tanh(\beta\omega/2)$ \cite{SMS89-FS91-KS856}. 
Here we chose to give the retarded polarizations. 
They are obtained by inserting 
$\pm i\, \epsilon \ {\rm sign}(p_0)$ in the denominator
of Eq. (\ref{propagator}).

All imaginary parts can be expressed in terms of three integrals
$E_1^{(i)}$, $E_2^{(i)}$, $E_3^{(i)}$ with $i \in \{ n,p\}$.
At finite temperature, for the calculation of the retarded 
polarizations, we have
\par\noindent {\bf --} for spacelike momentum:
\beq
E_n^{(i)} &=&  \int dy \left[ \left( y+{\omega \over 2} \right)^{n-1} 
\theta(y-y^{(i)}_L) \Bigl\{ n_i(y) -n_i(y+\omega) \Bigr\} \right. 
\nonumber \\
          & & \left. \qquad +(-1)^n
\left( y-{\omega \over 2} \right)^{n-1} \theta(y-y^{(i)}_U) 
\left\{ \overline n_i(y) - \overline n_i(y-\omega) \right\} \right] 
\eeq 
\par\noindent {\bf --} for timelike momentum with $q^2 < 4 M_i^2$,
the imaginary parts vanish: $E_n^{(i)} = 0$, 
\par\noindent {\bf --} for timelike momentum with $q^2 > 4 M_i^2$:
\beq
E_n^{(i)} &=& (-1)^{n-1} \int_{M_i}^\infty  dy 
\left( y -{\omega \over 2} \right)^{n-1} 
\left[ \theta(y-y_L^{(i)}) - \theta(y-y_U^{(i)}) \right]  
\left\{ n^{(i)} (\omega -y) + \overline n^{(i)} (y) -1 \right\}
\eeq
with
\beq
&& n_i(y)=\left[ e^{\beta (y-\mu_i)} +1 \right]^{-1} \ , \quad
\overline n_i(y)=\left[ e^{\beta (y+\mu_i)} +1 \right]^{-1} 
\nonumber \\
&& y_L^{(i)} = \left| {k \sqrt{\Delta^{(i)}} - \omega \over 2} \right|  
\ , \quad y_U^{(i)} = \left| {k \sqrt{\Delta^{(i)}} + \omega \over 2} 
\right|  \ , \quad \Delta^{(i)}= 1 - 4 {M_i^2 \over q^2} \nonumber
\eeq
\beq
{\cal I}m\ \Pi_\sigma^{(i)} &=& - {g_\sigma^2 \over 2 \pi k} 
\left( M_i^2 -{q^2 \over 4} \right) E_1^{(i)} \nonumber \\
{\cal I}m\ \Pi_{\omega\, L}^{(i)} &=&  {g_\omega^2 \over 2 \pi k} 
\left[ {q^2 \over 4} E_1^{(i)} - {q^2 \over k^2} E_3^{(i)} \right]  
\nonumber \\
{\cal I}m\ \Pi_{\omega\, T}^{(i)} &=&  {g_\omega^2 \over 4 \pi k} 
\left[ \left( M_i^2 + {q^2 \over 4} \right) E_1^{(i)} + 
{q^2 \over k^2} E_3^{(i)} \right] \nonumber \\
{\cal I}m\ \Pi_{\sigma\omega}^{(i)} &=& - {g_\sigma g_\omega 
\over 2 \pi k^3 }\ q^2 M_i\ E_2^{(i)} \nonumber \\
{\cal I}m\ \Pi_\delta^{(i)} &=& - {g_\delta^2 \over 2 \pi k} 
\left( M_i^2 -{q^2 \over 4} \right) E_1^{(i)} \nonumber \\
{\cal I}m\ \Pi_{\rho\, L}^{(i)} &=&  {1\over 2 \pi k} 
\left\{ g_\rho^2 \left[ {q^2 \over 4} E_1^{(i)} - {q^2 \over k^2} 
E_3^{(i)} \right]  + g_\rho {f_\rho \over 2 m} \ q^2 M_i \
E_1^{(i)} + \left(  {f_\rho \over 2 m} \right)^2 \left[ M_i^2 q^2\
E_1^{(i)} +{q^4 \over k^2} E_3^{(i)} \right] \right\}
\nonumber \\
{\cal I}m\ \Pi_{\rho\, T}^{(i)} &=&  {1 \over 4 \pi k} \left\{ 
g_\rho^2 \left[ \left( M_i^2 + {q^2 \over 4} \right) E_1^{(i)} + 
{q^2 \over k^2} E_3^{(i)} \right] 
+ 2 g_\rho {f_\rho \over 2 m} \ q^2 M_i \
E_1^{(i)}  \right. \nonumber \\
& & \left. \qquad + \left(  {f_\rho \over 2 m} \right)^2 \left[
\left( M_i^2 q^2 + {q^4 \over 4} \right) E_1^{(i)} 
-{q^4 \over k^2} E_3^{(i)} \right] \right\}
\nonumber \\
{\cal I}m\ \Pi_{\delta\rho}^{(i)} &=& - {g_\delta \left( 2 M_i\ 
g_\rho + (f_\rho/2 m) q^2 \right) q^2 
\over 4 \pi k^3 }\  E_2^{(i)} \nonumber \\
{\cal I}m\ \Pi_{\sigma\rho}^{(i)} &=&
 - {g_\sigma \left( 2 M_i\  g_\rho + (f_\rho/2 m) q^2 \right) q^2 
\over 4 \pi k^3 }\  E_2^{(i)} \nonumber \\
{\cal I}m\ \Pi_{\delta\omega}^{(i)} &=& - {g_\delta g_\omega \over 
2 \pi k^3 }\  M_i\ q^2 \ E_2^{(i)} \nonumber \\
{\cal I}m\ \Pi_{\sigma\delta}^{(i)} &=& - {g_\sigma g_\delta \over 2 \pi k} 
\left( M_i^2 -{q^2 \over 4} \right) E_1^{(i)} \nonumber \\
{\cal I}m\ \Pi_{\rho\omega\, L}^{(i)} &=&  {1\over 4 \pi k} 
\left\{ g_\rho g_\omega \left[ {q^2 \over 2} E_1^{(i)} - 2 {q^2 \over k^2} 
E_3^{(i)} \right]  + g_\omega {f_\rho \over 2 m} \ q^2 M_i \
E_1^{(i)}  \right\}
\nonumber \\
{\cal I}m\ \Pi_{\rho\, T}^{(i)} &=&  {1 \over 4 \pi k} \left\{ 
g_\rho g_\omega \left[ \left( M_i^2 + {q^2 \over 4} \right) E_1^{(i)} + 
{q^2 \over k^2} E_3^{(i)} \right] 
+  g_\omega {f_\rho \over 2 m} \ q^2 M_i \ E_1^{(i)} \right\}
\nonumber 
\eeq

\newpage

\begin{figure}[htb]
\mbox{%
\parbox{8.2cm}{\epsfig{file=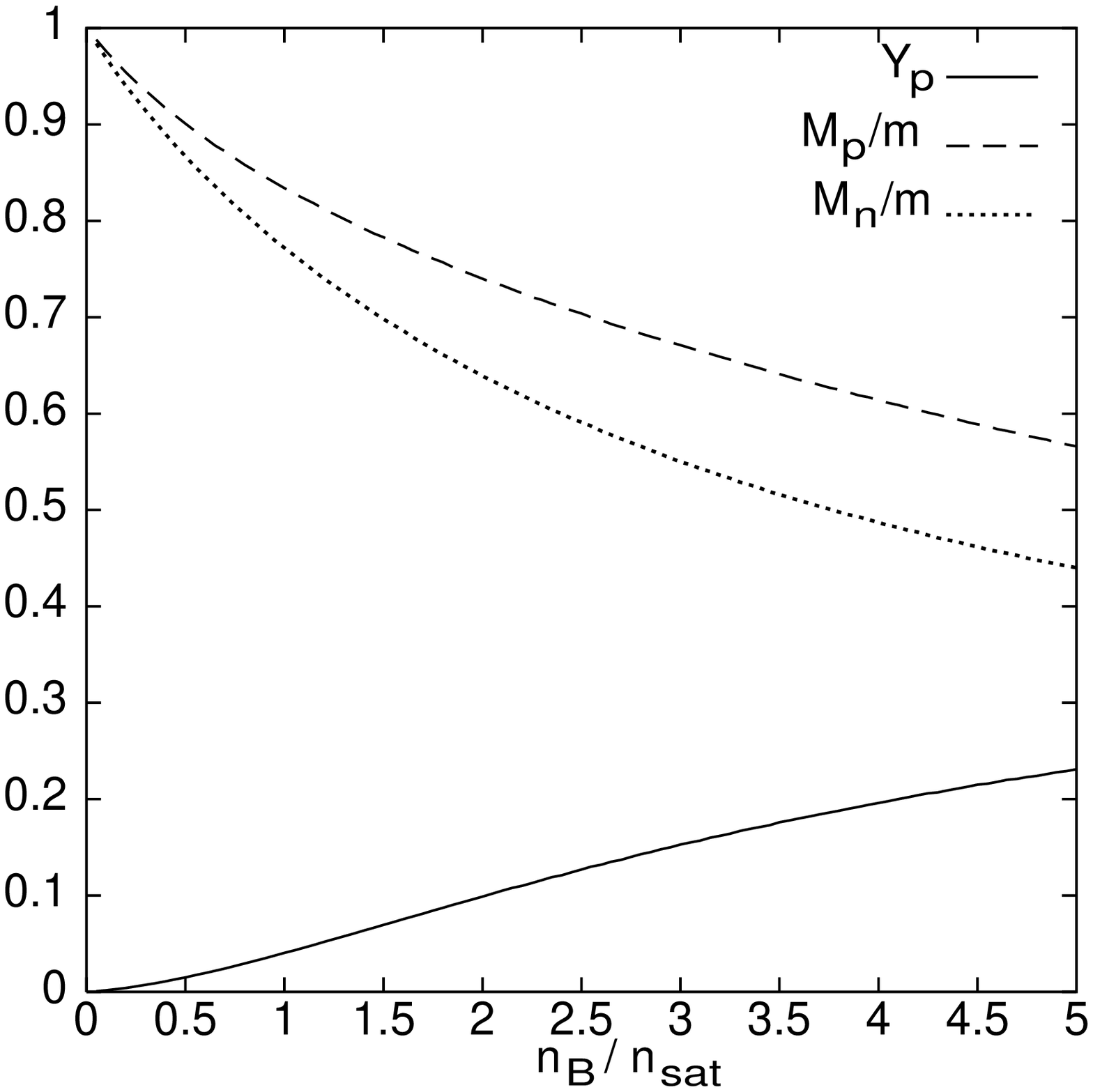,width=8.2cm}}
\parbox{0.2cm}{\phantom{a}}
\parbox{5.2cm}{\small {\bf Fig. 1} Effective masses and proton fraction 
as calculated from parameter set (\ref{fiteos}) with $g_\delta=5$
in neutrino-free matter in $\beta$ equilibrium at $T$=0.}
}
\end{figure}

\begin{figure}[htb]
\mbox{%
\parbox{7.2cm}{\epsfig{file=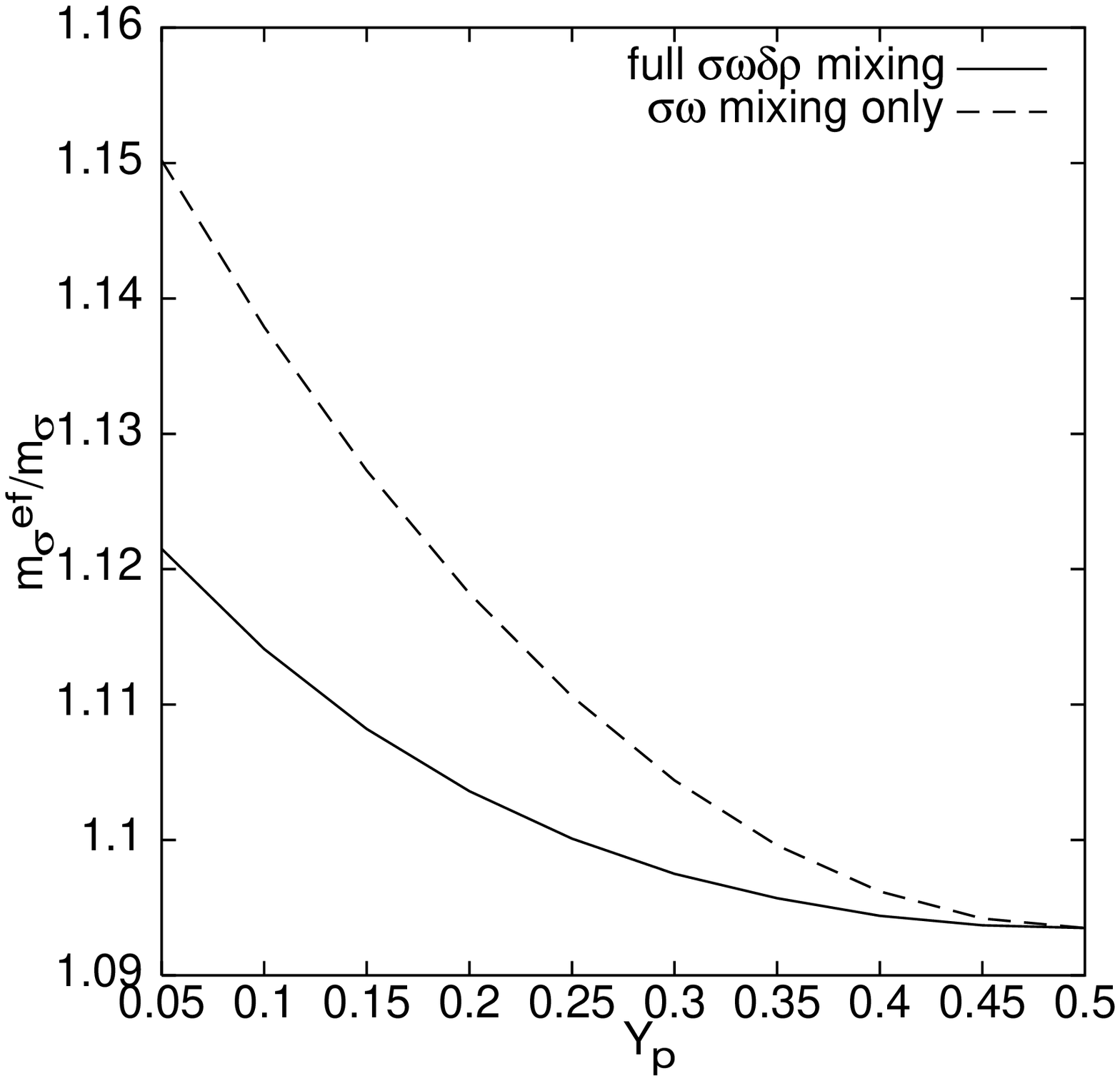,width=8.2cm}}
\parbox{0.2cm}{\phantom{aa}}
\parbox{7.2cm}{\epsfig{file=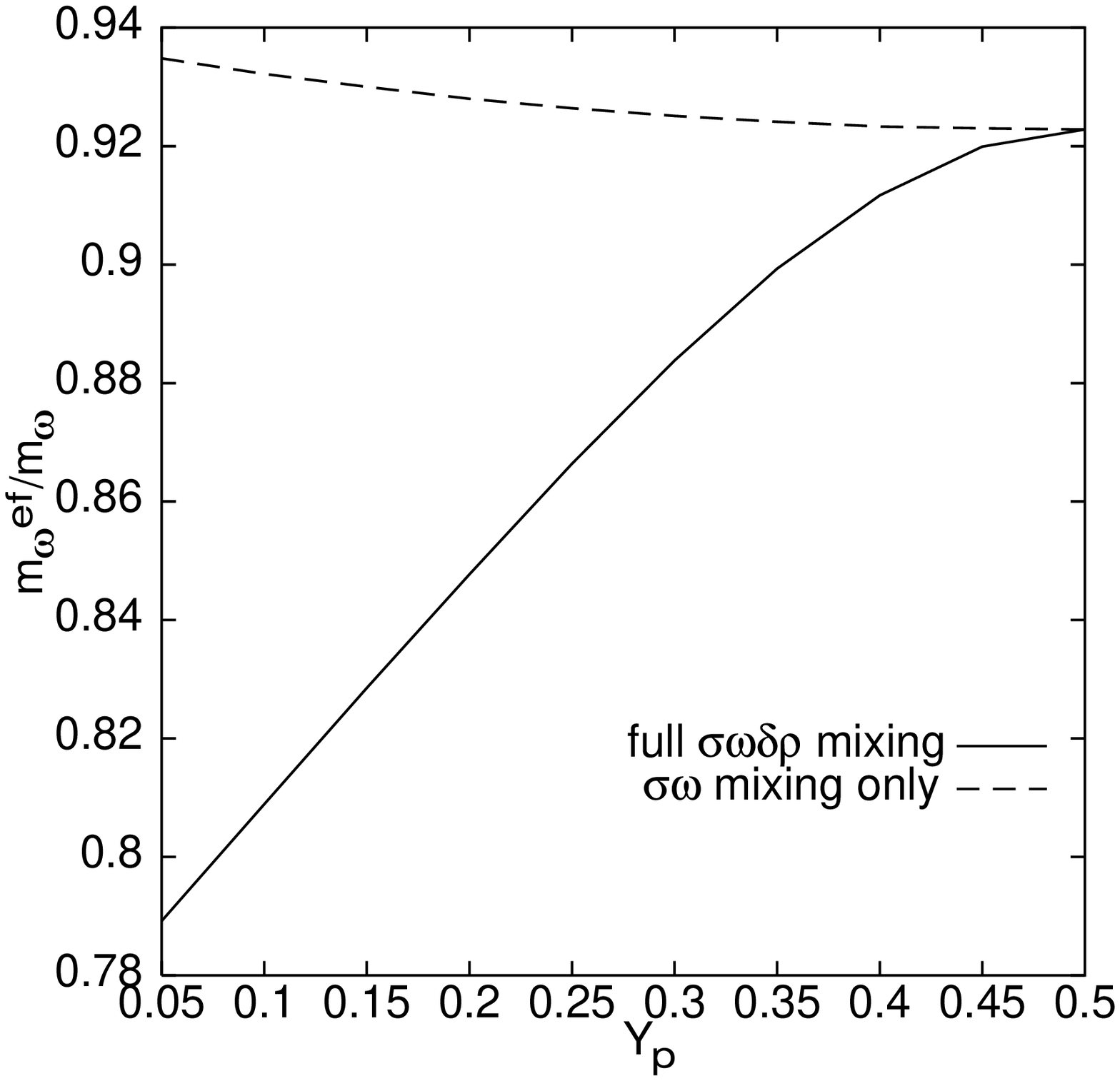,width=8.2cm}}
}
\mbox{%
\parbox{7.2cm}{\epsfig{file=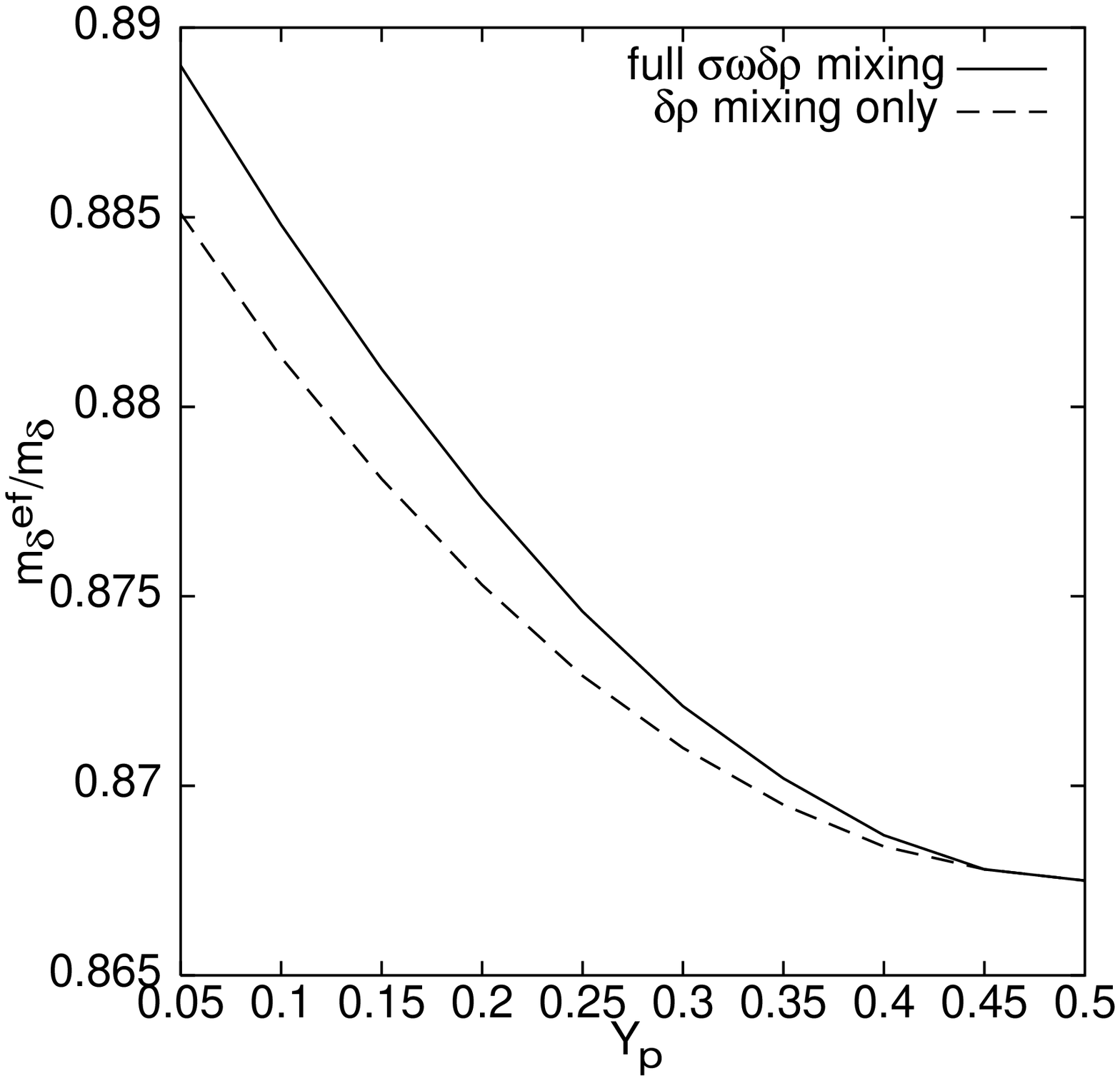,width=8.2cm}}
\parbox{0.2cm}{\phantom{aa}}
\parbox{7.2cm}{\epsfig{file=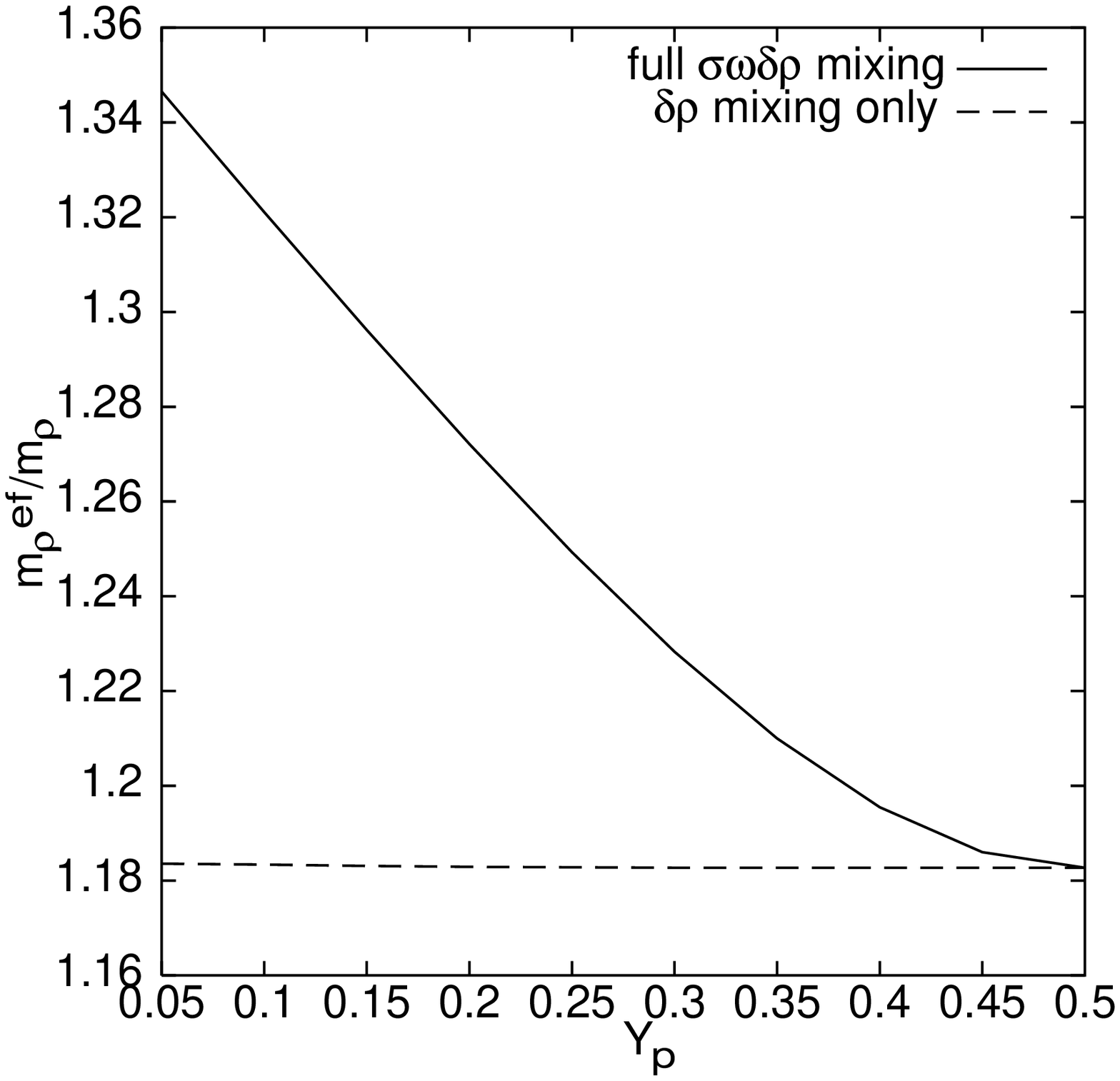,width=8.2cm}}
}
\mbox{%
\parbox{15cm}{\small {\bf Fig. 2} Effective masses of mesons as
calculated with renormalization procedure A as a function of 
asymmetry. The density was fixed at $n_B=4\, n_{\rm sat}$
and the temperature at $T=0$, and the coupling constants 
are those of (\ref{fiteos}). }
}
\end{figure}

\begin{figure}[htb]
\mbox{%
\parbox{7.2cm}{\epsfig{file=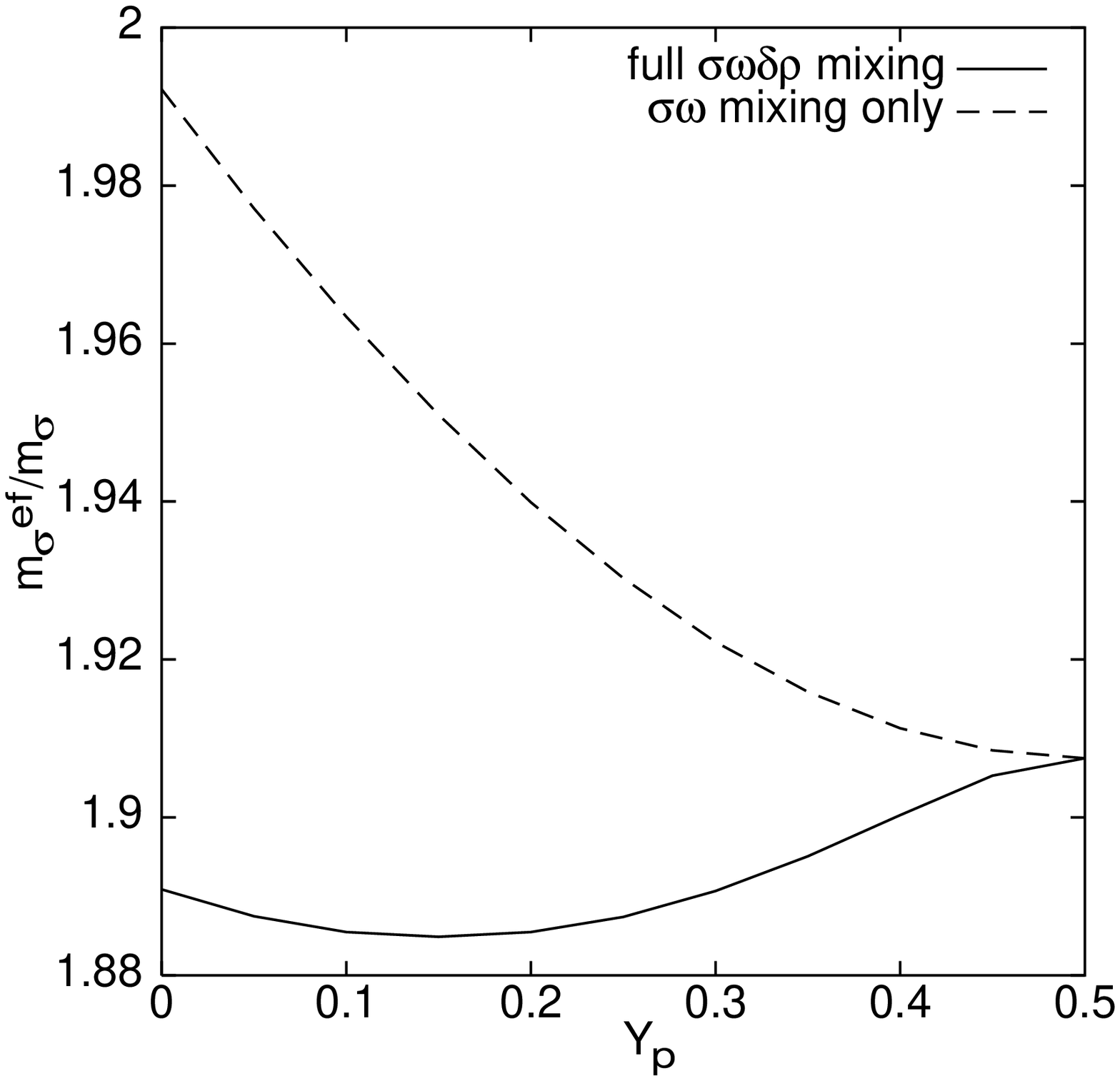,width=8.2cm}}
\parbox{0.2cm}{\phantom{aa}}
\parbox{7.2cm}{\epsfig{file=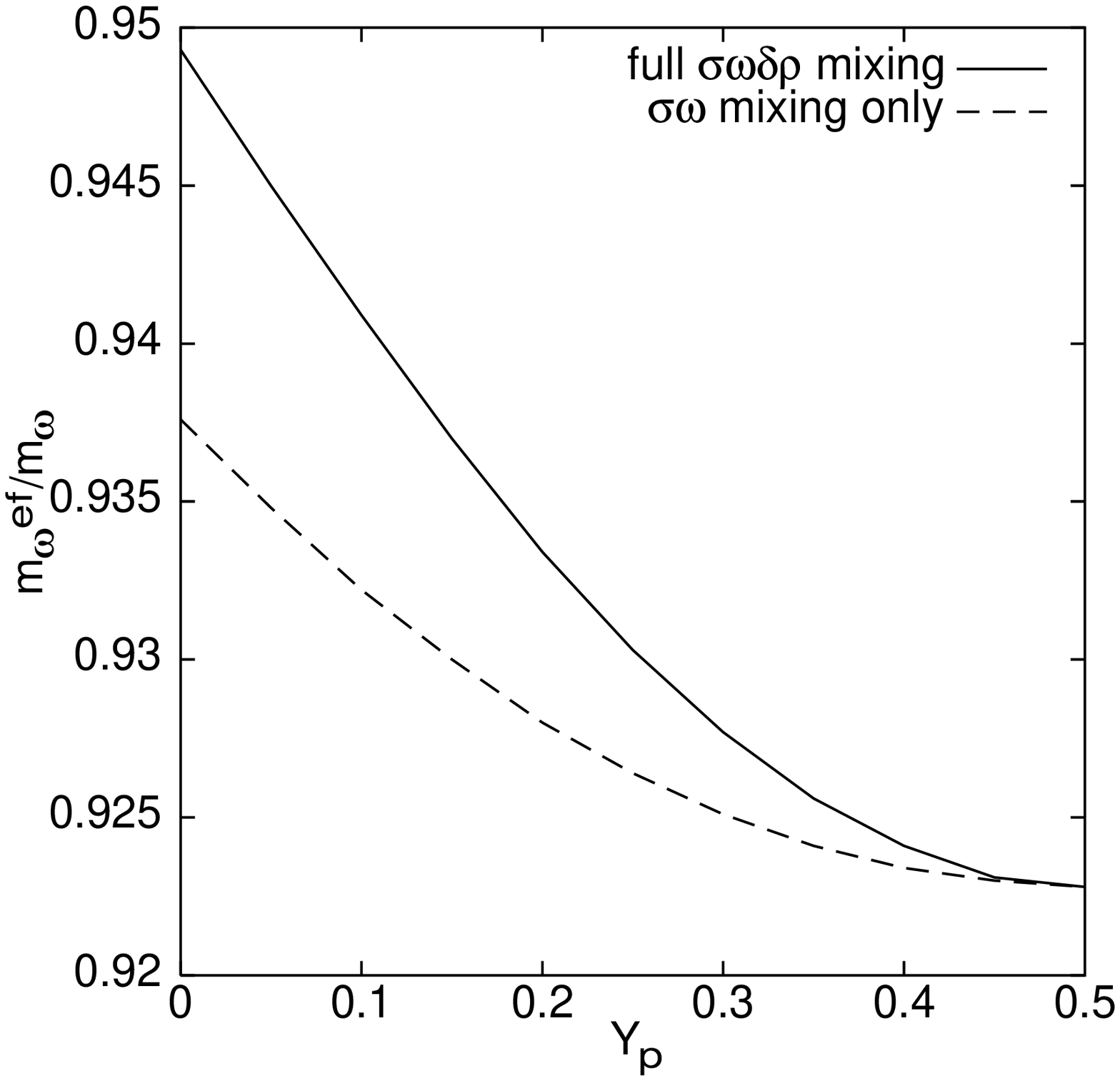,width=8.2cm}}
}
\mbox{%
\parbox{7.2cm}{\epsfig{file=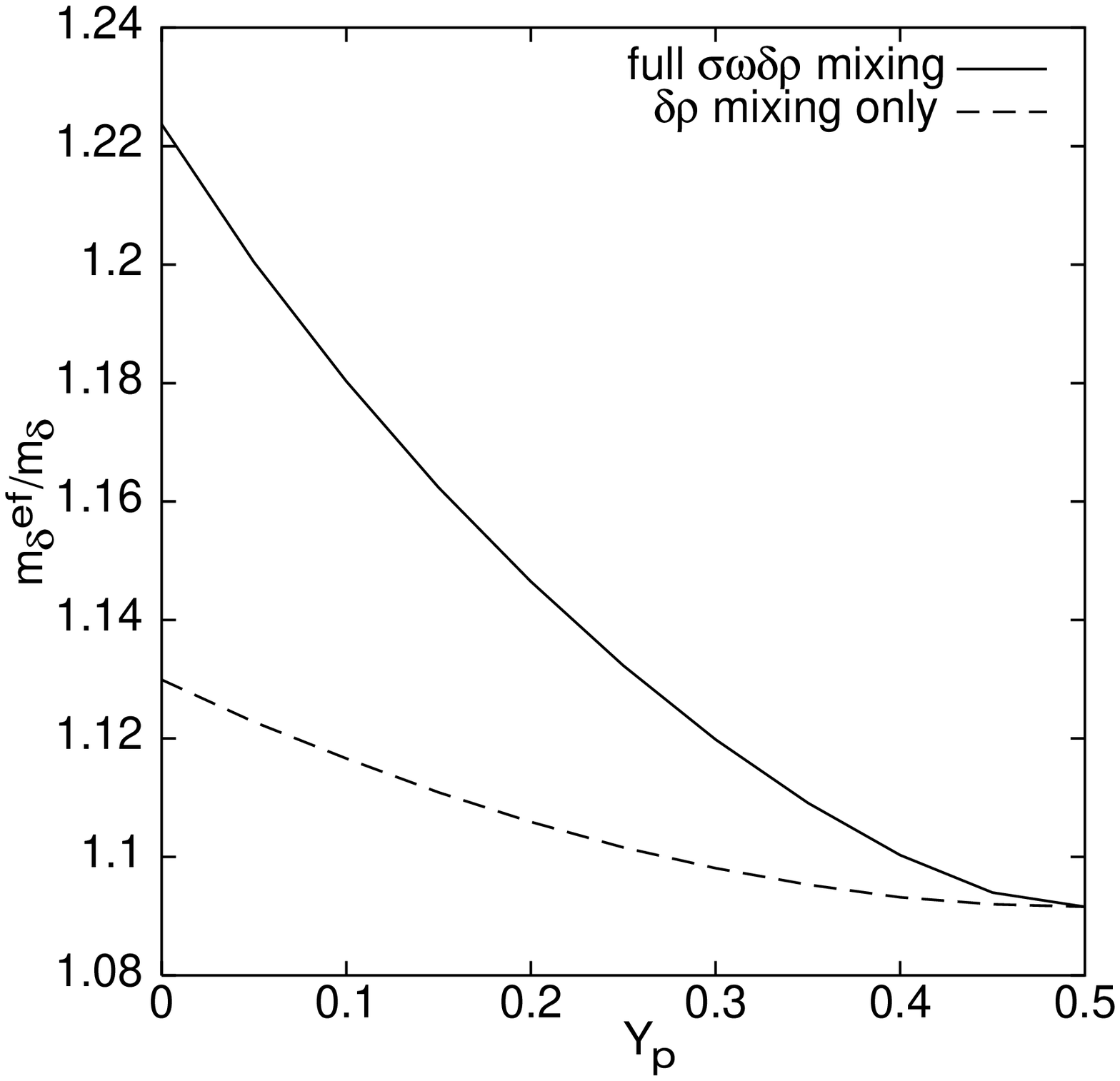,width=8.2cm}}
\parbox{0.2cm}{\phantom{aa}}
\parbox{7.2cm}{\epsfig{file=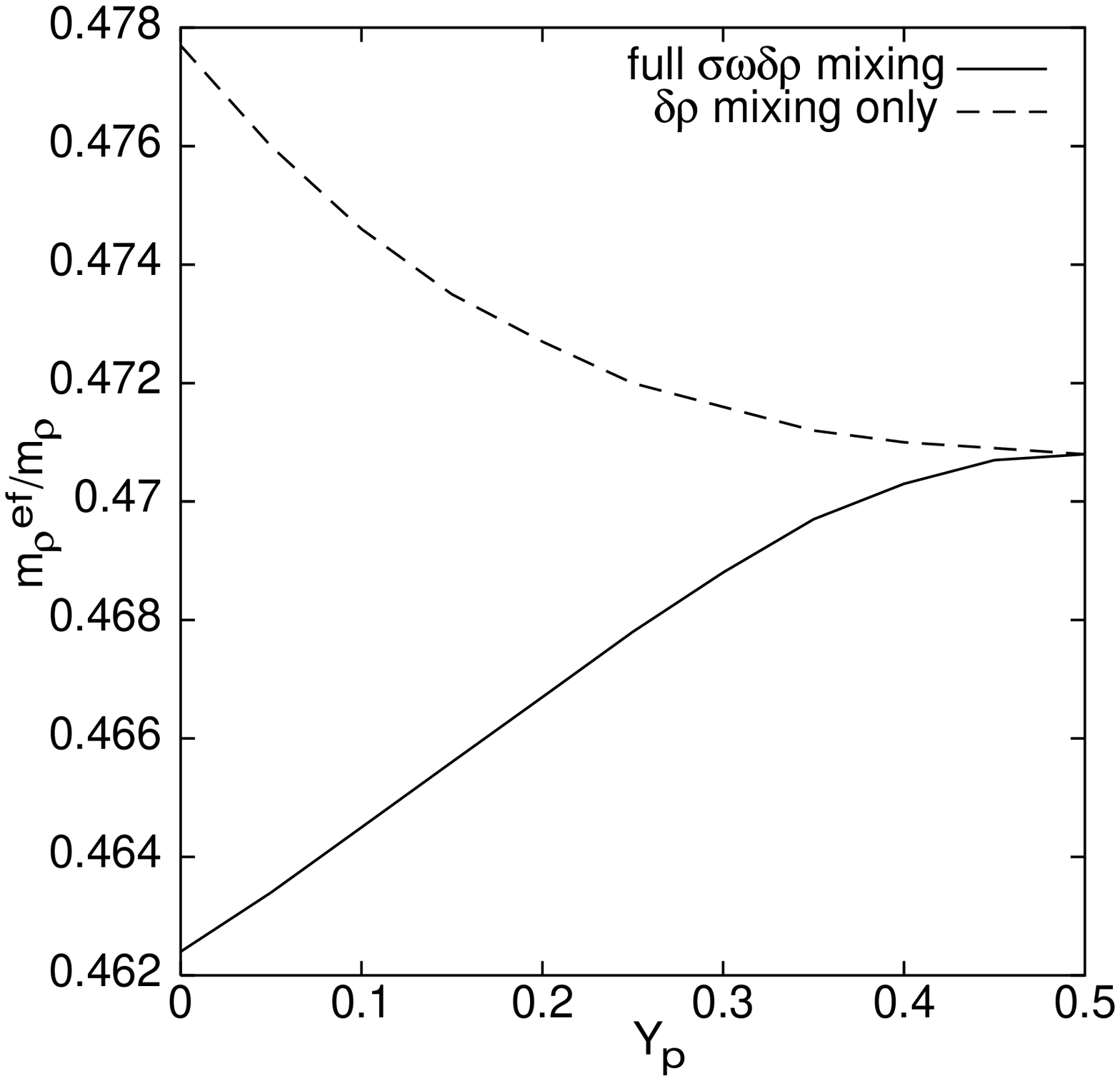,width=8.2cm}}
}
\mbox{%
\parbox{15cm}{\small {\bf Fig. 3} Effective masses of mesons as
calculated with renormalization procedure B as a function of 
asymmetry. The density was fixed at $n_B=4\, n_{\rm sat}$
and the temperature at $T=0$, and the coupling constants 
are those of (\ref{fiteos})}
}
\end{figure}

\begin{figure}[htb]
\mbox{%
\parbox{7.2cm}{\epsfig{file=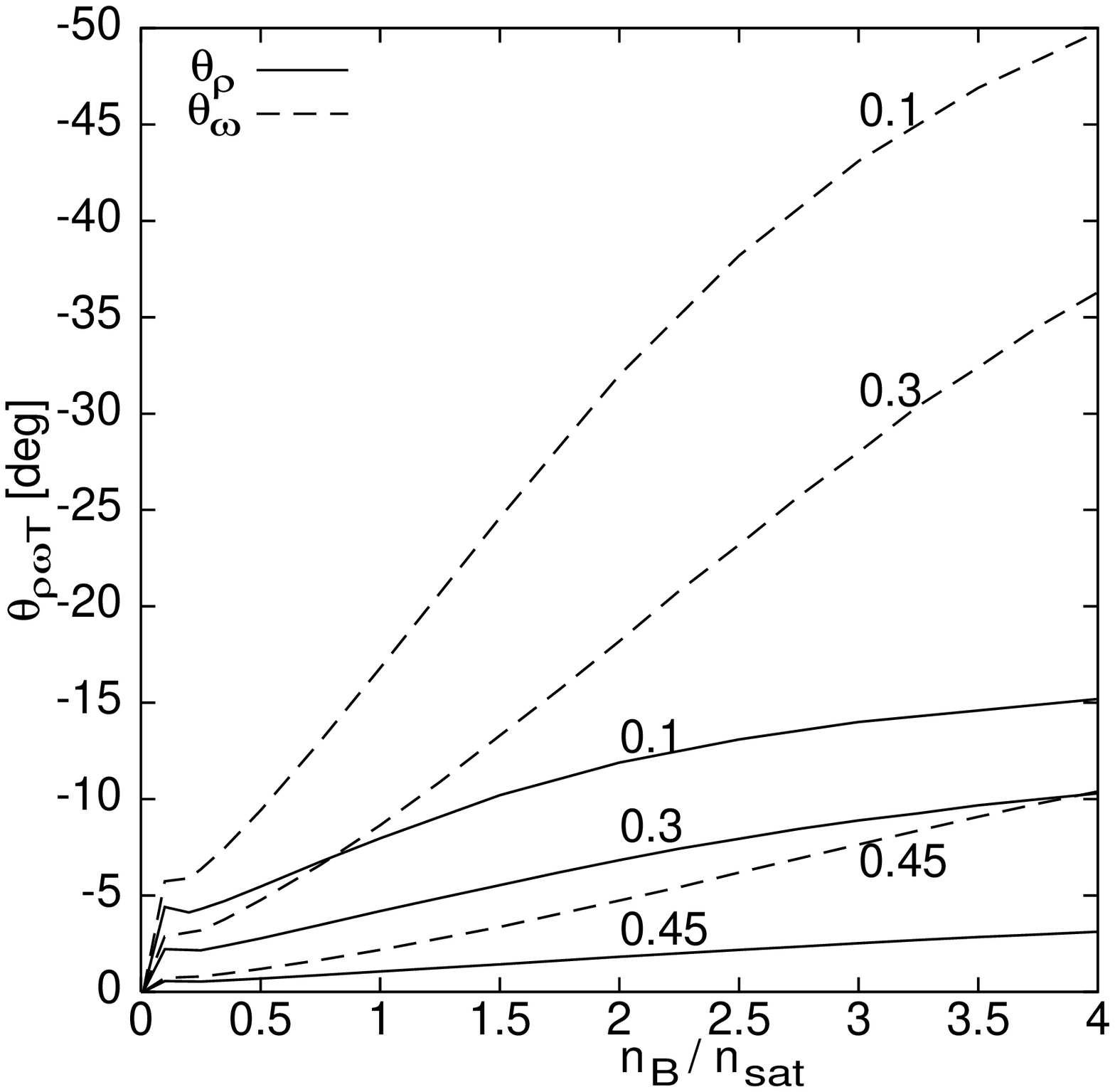,width=8.2cm}}
\parbox{0.2cm}{\phantom{aa}}
\parbox{7.2cm}{\epsfig{file=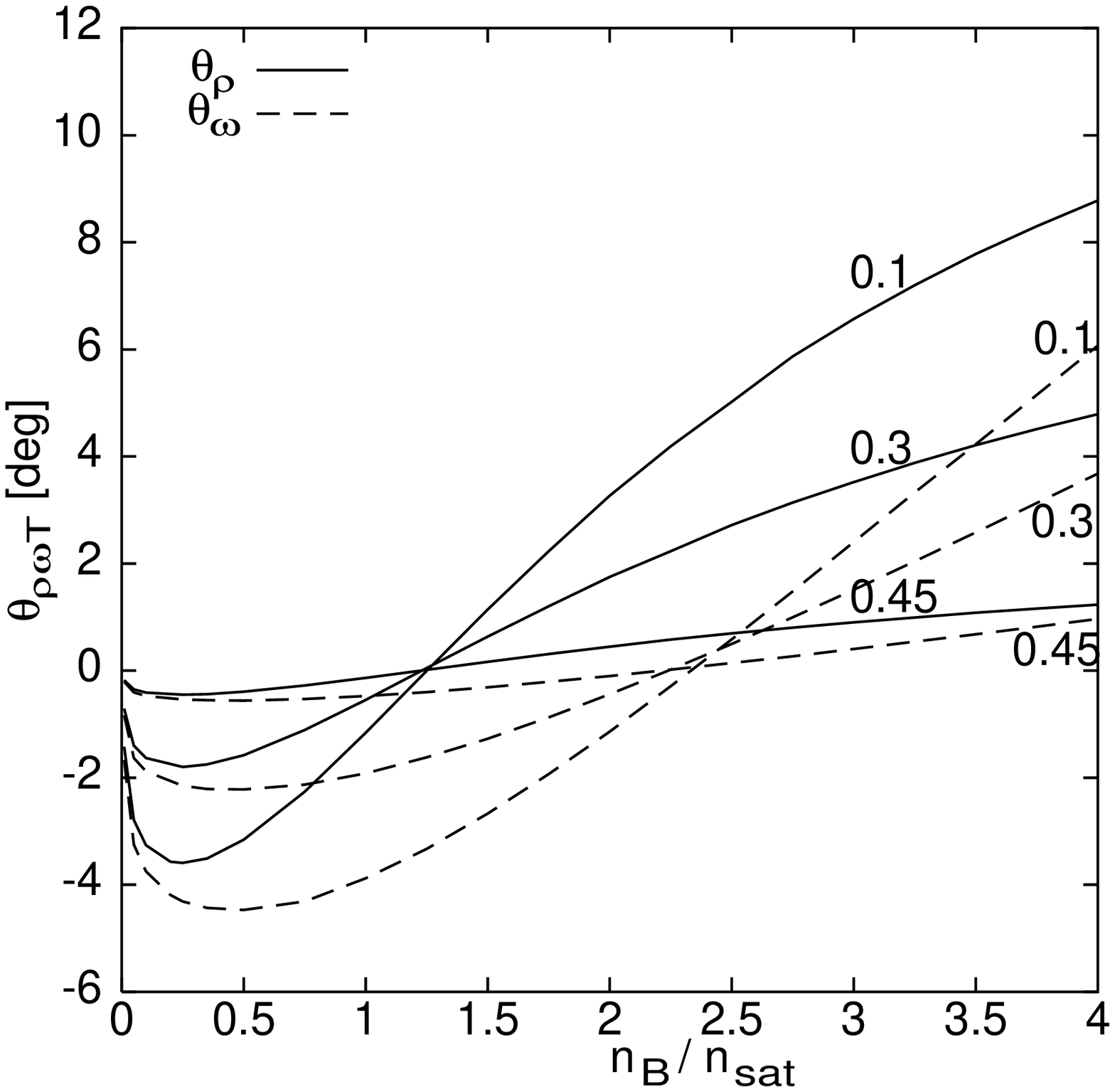,width=8.2cm}}
}
\vskip 0.5cm
\mbox{%
\parbox{15cm}{\small {\bf Fig. 4} $\rho$-$\omega$ mixing angle in the
transverse mode as a function of density asymmetry, as calculated with 
renormalization procedures A (left) or B (right). The parameters 
are those of \cite{M87} and the temperature is set to $T$=0. The
exchanged 3-momentum was fixed to $k=500$ MeV. The curves are labelled 
by the value of $Y_p$.}
}
\end{figure}

\begin{figure}[htb]
\mbox{%
\parbox{8.2cm}{\epsfig{file=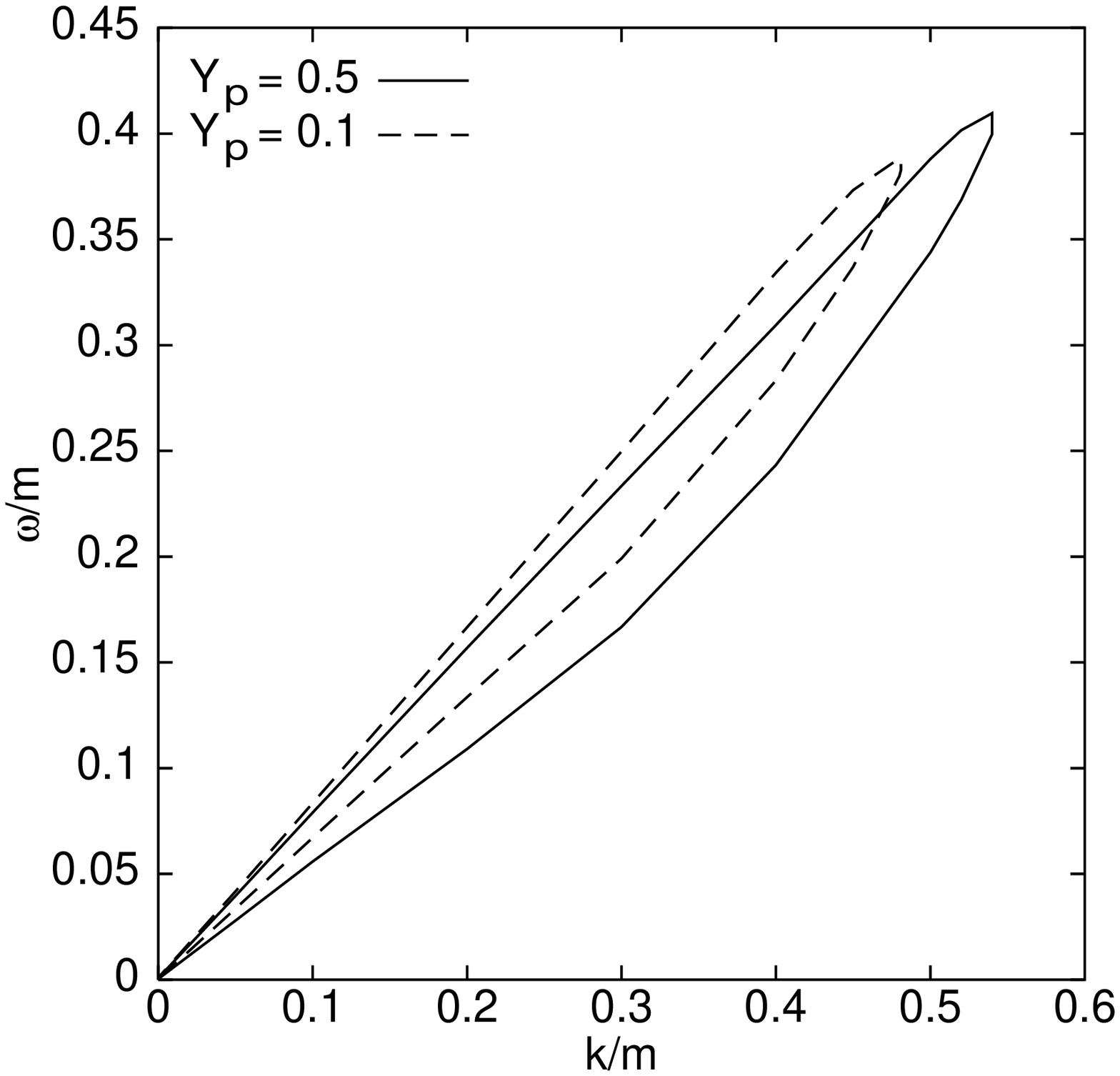,width=8.2cm}}
\parbox{0.2cm}{\phantom{aa}}
\parbox{6cm}{\small {\bf Fig. 5} Effect of asymmetry on the zero 
sound mode due to $\sigma$-$\omega$ mixing. The figure was plotted
for $n_B=4\, n_{\rm sat}$ and $T=0$ and using renormalization scheme (A). 
The coupling constants are those of \cite{M87}. }
}
\end{figure}

\begin{figure}[htb]
\mbox{%
\parbox{7.6cm}{\epsfig{file=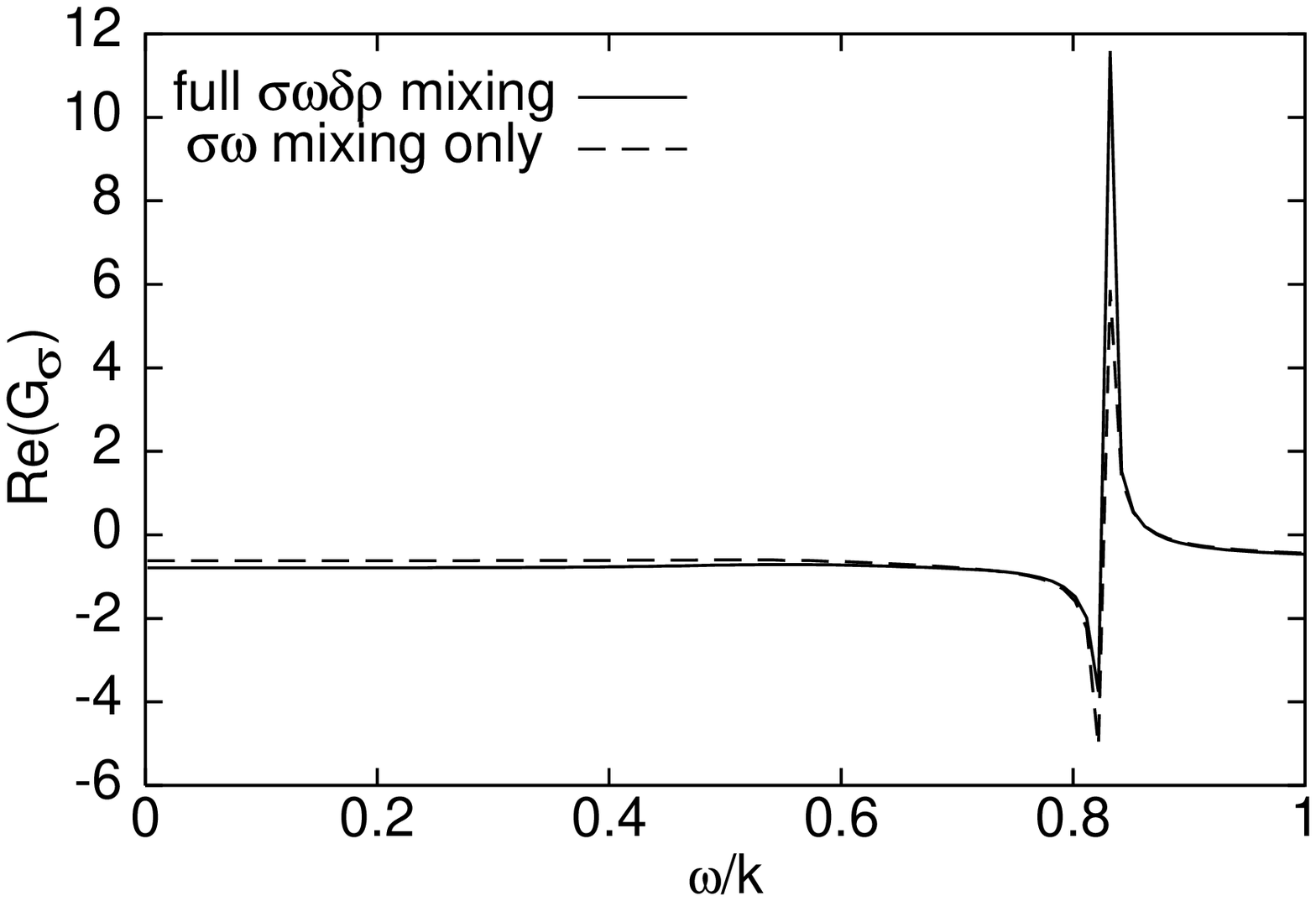,width=7.6cm}}
\parbox{0.2cm}{\phantom{aa}}
\parbox{7.6cm}{\epsfig{file=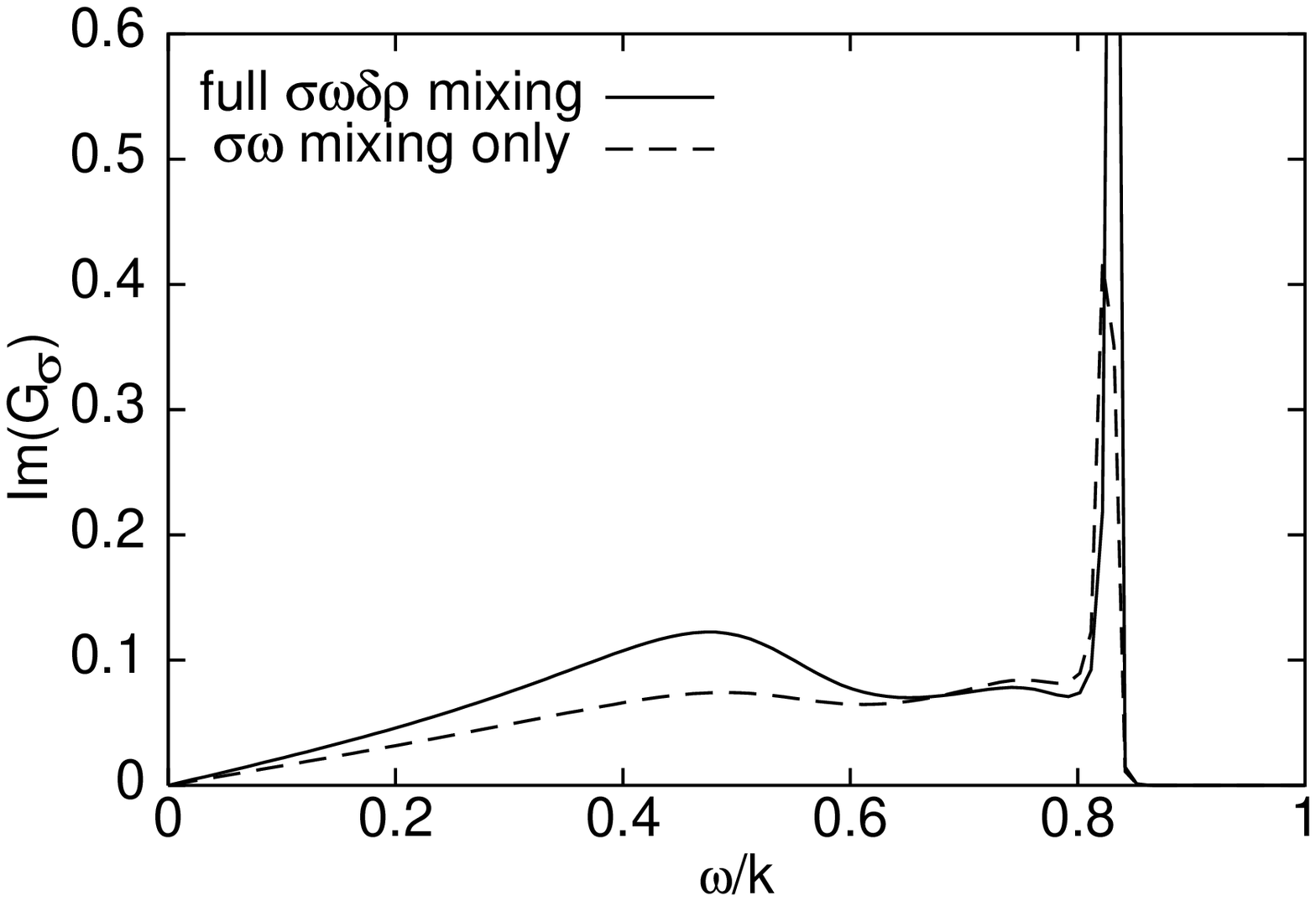,width=7.6cm}}
}
\mbox{%
\parbox{15.4cm}{\small {\bf Fig. 6} Real and imaginary parts 
of the $\sigma$ meson propagator.}
}
\end{figure}

\begin{figure}[htb]
\mbox{%
\parbox{7.6cm}{\epsfig{file=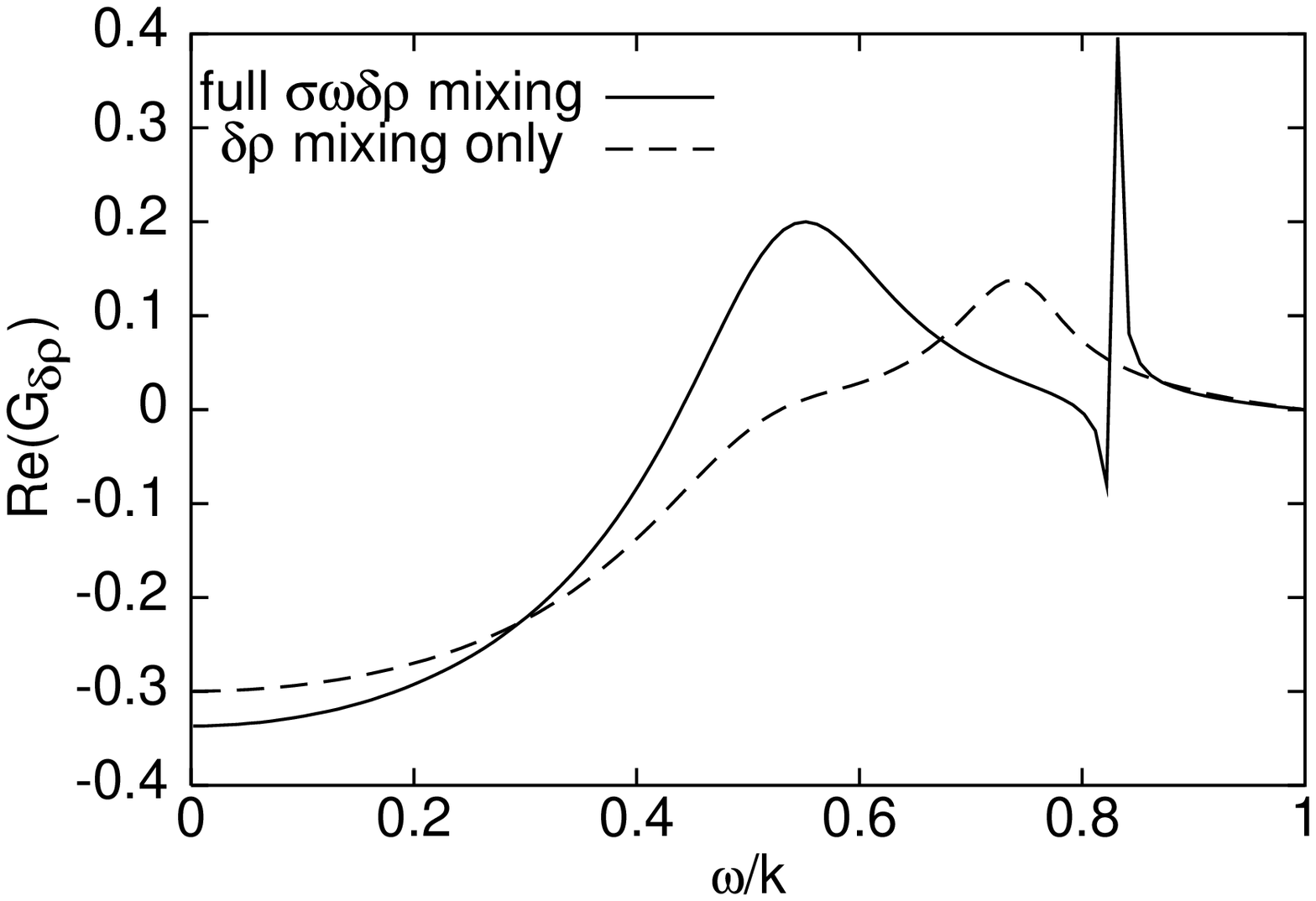,width=7.6cm}}
\parbox{0.2cm}{\phantom{aa}}
\parbox{7.6cm}{\epsfig{file=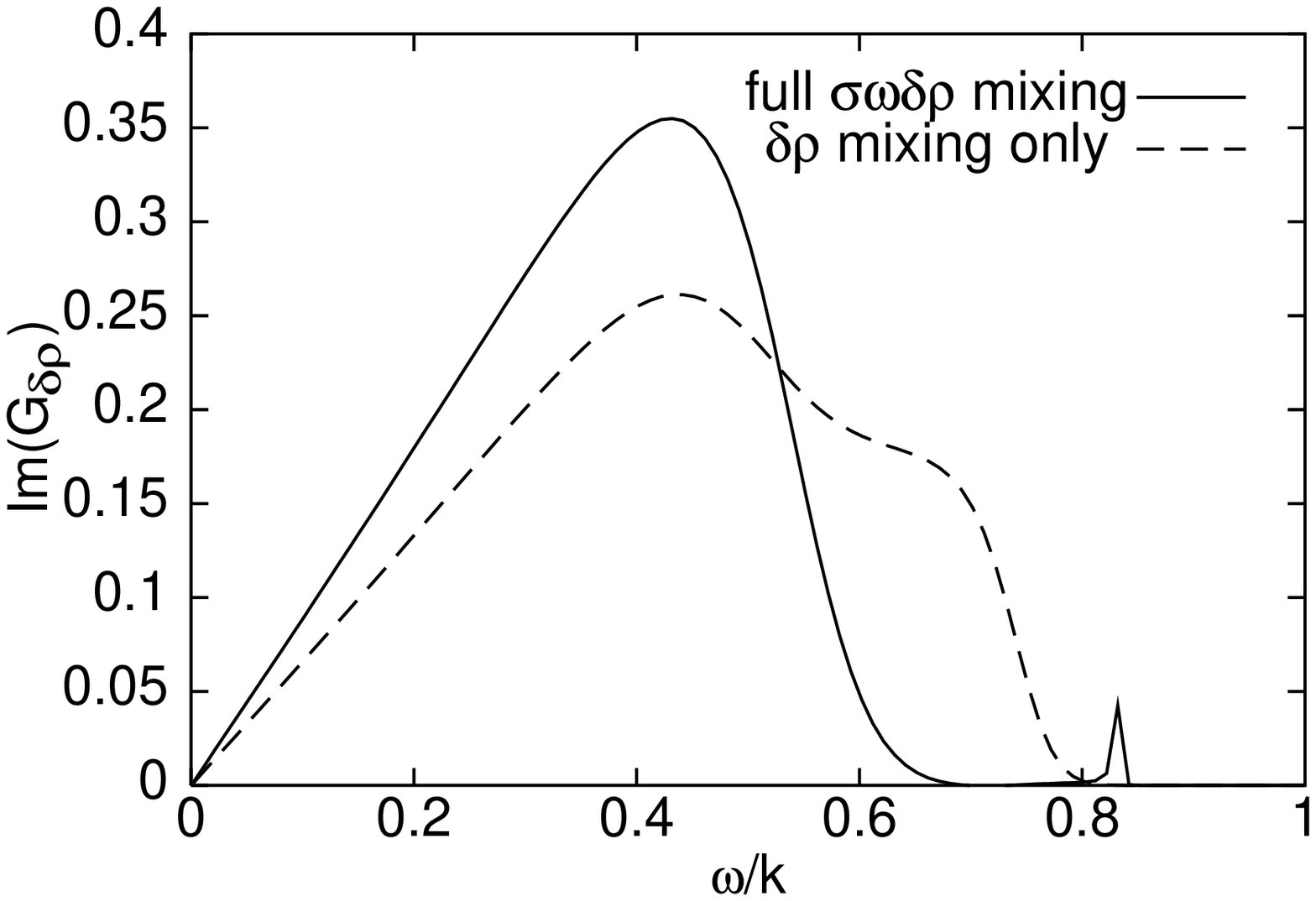,width=7.6cm}}
}
\mbox{%
\parbox{15.4cm}{\small {\bf Fig. 7} Real and imaginary parts 
of the mixed $\delta$-$\rho$ meson propagator.}
}
\end{figure}

\begin{figure}[htb]
\mbox{%
\parbox{7.6cm}{\epsfig{file=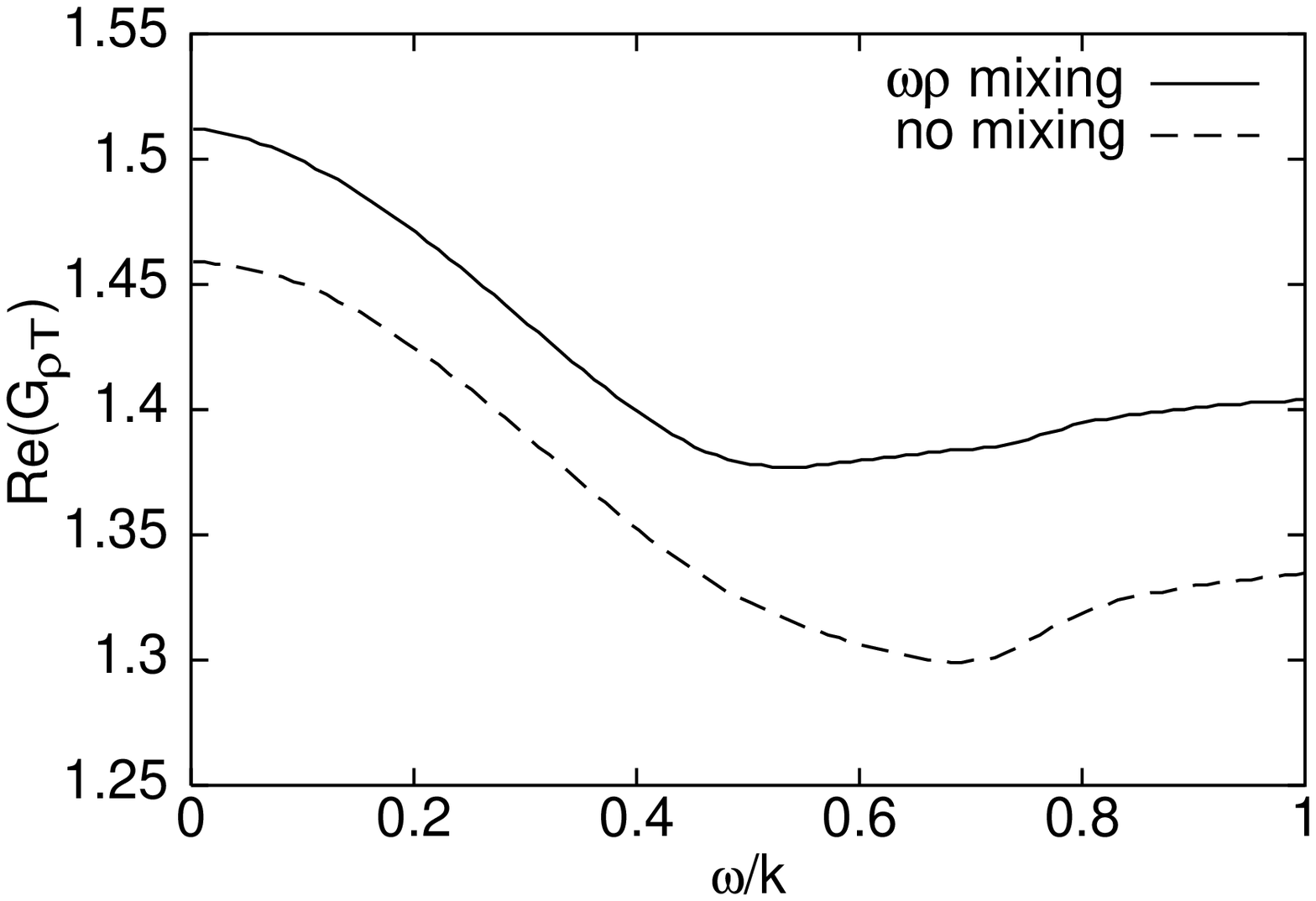,width=7.6cm}}
\parbox{0.2cm}{\phantom{aa}}
\parbox{7.6cm}{\epsfig{file=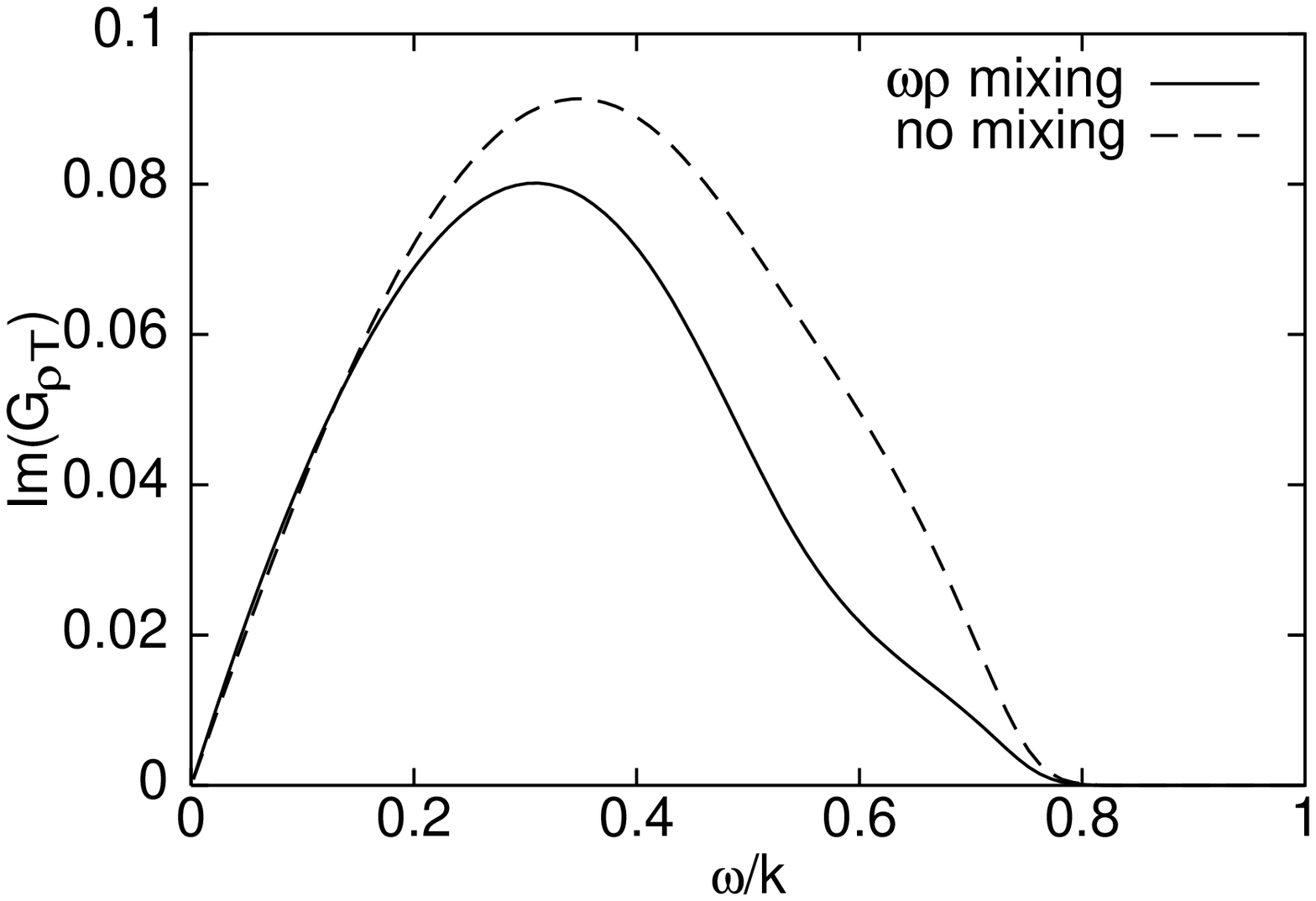,width=7.6cm}}
}
\mbox{%
\parbox{15.4cm}{\small {\bf Fig. 8} Real and imaginary parts 
of the transverse $\rho$ meson propagator.}
}
\end{figure}

\end{document}